\documentstyle[bezier]{article}

\input{prepictex}
\input{pictex}
\input{postpictex}

\def\href#1#2{{#2}}
\def\beq{\begin{eqnarray}}
\def\eeq{\end{eqnarray}}
\def\beqn{\begin{eqnarray*}}
\def\eeqn{\end{eqnarray*}}
\def\Tr{\mathop{\rm Tr}}
\def\TTr{\mathop{\bf \rm Tr}}
\def\Res{\mathop{\rm Res}}
\def\det{\mathop{\rm det}}

\def\->{\mathop{\longrightarrow}}
\def\Theta{\mathop{\theta}}
\newcommand {\n}{\nonumber \\}
\renewcommand{\theequation}{\thesection.\arabic{equation}}
\begin{document}
  \begin{normalsize}
    \begin{flushright}
      \href{http://xxx.lanl.gov/abs/hep-th/9810004}{hep-th/9810004}
    \end{flushright}
  \end{normalsize}
\begin{center}
\vspace*{1cm}
{\Large D-brane Scattering in \\IIB String Theory and IIB Matrix Model}
\vspace{2cm}\\
Yoshihisa K{\sc itazawa}$^{}$\footnote
{E-mail address: kitazawa@th.phys.titech.ac.jp}{\sc and}
\href{http://131.112.122.96/}
{Hiroyuki T{\sc akata}}$^{}$\footnote
{E-mail address: takata@th.phys.titech.ac.jp, JSPS Research Fellow \\
 \hspace*{13pt} URL  \hspace*{3pt} address: 
\href{http://www.th.phys.titech.ac.jp/~takata/}
{http://www.th.phys.titech.ac.jp/\~{\hspace*{0pt}}takata/}
 }\\
\vspace{1cm}
\href{http://www.th.phys.titech.ac.jp/particle/particle.html}
{
{\it Department of Physics, Tokyo Institute of Technology,}\\
{\it Oh-okayama, Meguro-ku, Tokyo 152-8551, Japan}
}
\\
\end{center}
\setcounter{equation}{0}
\vfill
\begin{abstract}
We consider two Dirichlet $p$-branes with lower dimensional brane charges 
and their scattering. 
We first calculate the cylinder amplitude of open string with suitable boundary conditions. 
We compare this result with that in the IIB matrix model. 
We find the agreement between them in the long distance,
low velocity, or large field limit. 
We also find a way to investigate more general boundary conditions for open string.
\end{abstract}
\vfill
\section{Introduction}
\setcounter{equation}{0}
String theory is the best candidate for the unified theory of interactions. 
Dirichlet branes~\cite{P} on which fundamental open string boundaries are attached are key objects to understand non-perturbative effects of string theory.
That is, in the type IIB string theory, 
the D-string is a strong-weak coupling dual object of the fundamental string.  
Furthermore a matrix model for IIB string theory was presented by IKKT~\cite{IKKT}. 
They propose that the model is  a non-perturbative definition of string theory. 
However, clear correspondences between string theory and its matrix model are not found yet. 
Of course  in Ref~\cite{IKKT}, 
some correspondences between them were presented.
Firstly, Green Schwarz action in Schild gauge can be related to the action of the matrix model in the case of the infinite matrix size.
Next, the potential between  two static D-strings in the matrix model was calculated.
In Ref~\cite{M} the authors calculated the scattering amplitude of two D-strings based on the matrix model and they compared it with the result~\cite{B} in string theory.
They set the non-zero values of all commutators of the coordinate matrices to $\alpha^{\prime}\times \mbox{constant}$.
However in Ref~\cite{CT}, the authors insist that the value of the commutators in the matrix model $F^M$ is the inverse of the field strength $F$ on the D-brane world volume in string theory in the large $F$ approximation.
In this context, the authors in Ref~\cite{MM} compared the potentials between two static D-branes in string theory and the IIB matrix model.\\ 
We consider D$p$-branes with non-zero field strength in IIB string theory and 
in its matrix model (IKKT model).
We calculate the amplitude of the scattering of two D$p$-branes and compare it with the one~\cite{M} of the IIB matrix model.
Since the relative velocity ($2v$) is non zero, the configuration is not BPS.
When $v \rightarrow 0$,
 the amplitude is expected to vanish.  
We find that the amplitudes in the two models are equal 
in three independent limits
if we identify the respective field strengths suitably.
Namely, we find how the "D-brane" configurations in the matrix model are translated to those in string theory. \\
\hspace*{5pt}
In section 2, we calculate the amplitude of two D$p$-branes with the field strength on their world volumes in bosonic sector.
First, we explain the configuration of two D-branes in the flat space-time.
Two D$p$-branes are extended infinitely in the world volume direction. 
They have a constant relative velocity to each other and flat world volume.
They are located in spatially parallel to each other.
In this situation we consider the fundamental open string whose ends are attached on the two D-branes.
Then we calculate the one loop amplitude (i.e. cylinder amplitude) of open string. 
When the equations of motion are chosen to be free field type, 
the boundary condition of open string is complex because of the non-zero field strength $F$ and the non-zero relative velocity $2v$.
Previously, $v=0$ case was calculated in~\cite{MM, GG}, $F=0$ case was also done in~\cite{B} and related case in IIA string theory was presented in~\cite{L}. 
We formulate a general mode expansion of the string coordinates in the flat world volume case, quantize them canonically, 
and construct the Virasoro algebra.\\
\hspace*{5pt}
In section 3, we consider superstring based on the NSR formalism.
In this paper, the boundary condition for the world sheet fermion is defined by the super transformation of the bosonic one.
Then we perform a calculation using the proper time representation of the amplitude. 
To compare the result with that of the matrix model, 
we choose the leading term in 
three independent expansions in terms of parameters, 
in the long distance ($b \gg \alpha^\prime$) case
where $b$ is the distance between two D-branes, 
in the low velocity case, 
or in the large gauge field strength
 case.\\
\hspace*{5pt}
In section 4, we consider the amplitude in the IKKT matrix model with a brief review in our notations.\\
\hspace*{5pt}
In section 5, 
we compare the amplitudes in the two models and identify the field strength $F$ in string theory with the value $F^M$ of the commutator in the matrix model, 
in such a way that $\det{(\eta +F)F^{M}}=-1$ where $\eta$ is the target space metric.
Then we find that the amplitudes in the two models are  precisely the same 
when $b$ is large for arbitrary $v$, $F$,
$v$ is small for arbitrary $b$, $F$,
or $F$ is large for arbitrary $b$, $v$.
when the distance between the two D$p$-branes is large 
for arbitrary velocity $v$. \\
\hspace*{5pt}
In section 6, there are conclusions and discussions.
\hspace*{5pt}
In Appendices, we explain a general method to study the open string boundary condition which is any linear combinations of Neumann and Dirichlet boundary conditions. 
Some detailed calculations are also explained there.
\subsection{physical situation}
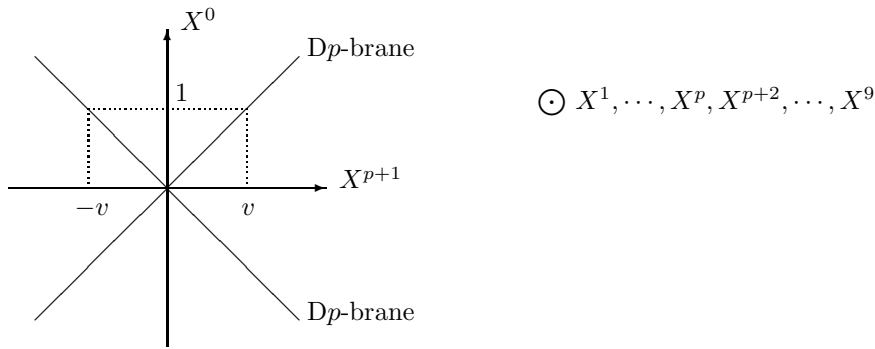
\begin{picture}(500,150)
\put(60,0){\vector(0,1){120}}
\put(0,60){\vector(1,0){120}}
\put(60,60){\line(1,1){50}}
\put(60,60){\line(-1,-1){50}}
\put(60,60){\line(1,-1){50}}
\put(60,60){\line(-1,1){50}}
\put(65,120){$X^0$}
\put(125,60){$X^{p+1}$}
\put(113,110){D$p$-brane}
\put(113,10){D$p$-brane}
\put(25,50){$-v$}
\put(88,50){$v$}
\put(63,93){1}
\bezier{30}(30,90)(60,90)(90,90)
\bezier{15}(30,90)(30,75)(30,60)
\bezier{15}(90,90)(90,75)(90,60)
\put(200,90){$\bigodot$ $X^1,\cdots, X^p,X^{p+2},\cdots,X^9$}
\end{picture}
First, we consider  two flat identical D-branes with the same world volume dimensions in the $D=10$ dimensional Minkowski space-time whose coordinates are written as $X^0,\cdots,X^{D-1}$.
We restrict the world volume dimension: $(p+1)$ to be even because we consider the IIB string theory.
We locate them spatially parallel to each other in the $D$ dimensional Minkowski space-time.
We call the common directions of the two D$p$-branes $1,\cdots p$, 
where  the world volumes of the D$p$-branes are infinitely extended.
We set two D$p$-branes to have relative velocity, $2v$, in the $(p+1)$-th direction. We set the coordinates' value of the $p+2,\cdots,(D-2) $-th directions of  the two D$p$-branes to be $0$ and locate them separately in the ($D-1$)-th direction by the distance $b$. 
We restrict $v$ and $b$ to be positive constant in this paper.
\subsection{correspondence between string theory and matrix model}
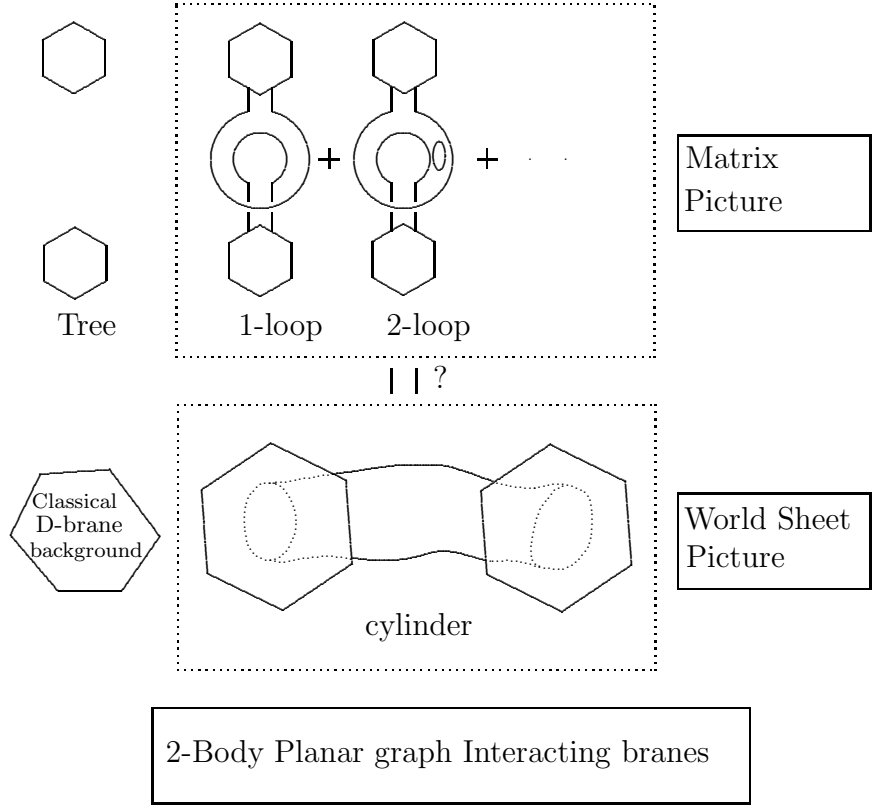
\input{fig2.pictex}
\vspace{.5cm}\\
Assuming that two D-branes interact by exchanging closed fundamental strings 
and any effect of two D-branes can be treated as open string boundary conditions, 
we calculate the amplitude of the fundamental string diagram 
with two boundaries, 
especially the cylinder amplitude. 
Since what is calculated in the matrix model is one loop diagram 
which is planar, 
we expect that the cylinder amplitude agrees with the sum of planar diagrams
in the IIB matrix model. 
The cylinder amplitude can be thought of the one loop amplitude of an open string whose boundaries are fixed in the D-branes, 
which we are considering now. 
\section{D$p$-brane-D$p$-brane scattering for bosonic part}
\setcounter{equation}{0}
We would like to calculate the scattering amplitude, 
that is the effective action. 
To calculate the amplitude, 
we have to obtain the propagators in space-time,  
in another word, 
the normal ordered Hamiltonian with the correct zero point energy.  
Therefore, 
in the following part in this section we define a normal ordering,  
construct Virasoro operators, 
and find a physical state condition.
Let us start with the following  world sheet action of open string.  
\beq 
S_{\rm bosonic}&=&-\frac{1}{4 \pi \alpha^{\prime}} \int d\sigma d\tau \partial_a X^\mu \partial^a X_\mu  \n
&+& \frac{1}{2\pi \alpha^\prime} \int d\tau A_{\mu}^0 \partial_\tau X^\mu \Bigg|_{\sigma=0}-\frac{1}{2\pi \alpha^\prime} \int d\tau A_{\mu}^{\pi} \partial_\tau X^\mu \Bigg|_{\sigma=\pi}\;,  \label{S_bosonic}
\eeq
where $A_{\mu}^{0}$ and $A_{\mu}^{\pi}$  are the gauge potentials on the D-brane world volume to which the  $\sigma=0 $ and $\sigma=\pi$ end of open string are attached, respectively. 
\footnote{In this paper, the indices run as follows: 
\beqn
\cases{ 
\mu,\nu,\rho,\sigma=0 \cdots D-1 & $\cdots$  in the D=10 dimensional space-time  \cr 
\alpha,\beta,\gamma=0,\cdots p   & $\cdots$  in the $p$-brane world volume  \cr  
\lambda,\delta=2,\cdots, D-1 & $\cdots$ in the pure Neumann and pure Dirichlet directions \cr 
a,b=0,1 & $\cdots$ the meaning depends on the situations \cr  
i,j = 2,\cdots ,p+1 &$\cdots$ in the 0 mode directions \cr 
k=p+2,\cdots D-1 & $\cdots$ in the pure Dirichlet direction \cr 
l,l^\prime = 1,\cdots, {p+1 \over 2} &$\cdots$ for the field strength \cr 
m,n \in {\bf Z} &$\cdots$ for bosonic modes and $b, c$ ghosts \cr 
r,s \in {\bf Z} \;\mbox{or}\; {\bf Z} +{1 \over 2} & $\cdots$ for fermionic modes and $\beta, \gamma$ ghosts  \cr 
} 
\;.
\eeqn
}
\\ 
We can rewrite the action as follows. 
\beqn
S_{\rm bosonic}&=&\left\{ 
\mbox{terms whose deformation vanishes with free equations of motion} 
\right\} \n 
&+& S_{\rm boundary} \;\;, \n
\eeqn 
where
\beq S_{\rm boundary} &=&-\frac{1}{4 \pi \alpha^\prime} X^\mu \left( \partial_\sigma X_\mu -2\partial_\tau A^{0}_{\mu} \right) \Bigg|_{\sigma=0} \n 
&+& \frac{1}{4 \pi \alpha^\prime} X^\mu \left( \partial_\sigma X_\mu -2\partial_\tau A^{\pi}_{\mu} \right) \Bigg|_{\sigma=\pi}\;. 
\eeq 
To get free equations of motion, we require  
\beq 
\delta S_{\rm boundary}=0\;. 
\eeq 
Namely, 
\beq 
&& \delta X^\mu \left. \left( \partial_{\sigma} X_{\mu} -2\partial_{\tau} A^{0}_{\mu} \right) \right|_{\sigma=0} =0 \;, \n
&& \delta X^\mu \left. \left( \partial_\sigma X_\mu -2\partial_\tau A^{\pi}_{\mu} \right) \right|_{\sigma=\pi} =0\;. 
\eeq 
\subsection{$p=1$ case }
In this subsection, we consider $p=1$ case to understand how to treat the non trivial boundary conditions.\\ 
First, imagine a pure (with no gauge field) D-string moving in the 2-nd direction with the constant velocity $v$. Its coordinates are restricted as
\beq
X^2 = v X^0 \;, 
\eeq
it is equivalent to
\beq
\partial_\tau \left( X^2-v X^0 \right)=0 \;.
\eeq
This is a Dirichlet boundary condition.\\
Second, in order to introduce gauge field on the world sheet of the D-string,
look at it from the proper time coordinate: $\widetilde{X}$ in the following way.
\beq
\partial_a X^\mu \longrightarrow \partial_a \widetilde{X^\mu}=\Lambda^{\mu}_{v\;\nu} \partial_a X^\nu \;, 
\eeq
so that 
\beq
\partial_\tau \widetilde{X}^2=0 \;,
\eeq
where
\beq
{\Lambda_v}^\mu_{\;\nu}=\left( 
\begin{array}{cc}
\begin{array}{ccc}
 \frac{1}{\sqrt{1-v^2}}& 0 & \frac{-v}{\sqrt{1-v^2}}\\
0 & 1 & 0\\
\frac{-v}{\sqrt{1-v^2}} & 0 &\frac{1}{\sqrt{1-v^2}}
\end{array}
&
{\bf 0}\\
{\bf 0}&
{\bf 1}
\end{array}
\right)^\mu_{\;\nu}\;.
\eeq
The boundary condition in this coordinate is, because of Lorentz invariance,  
\beq
\delta \widetilde{X}^\mu \left( \partial_{\sigma} \widetilde{X}_{\mu} -2\partial_{\tau} \widetilde{A}_{\mu}^{0, \pi} \right) =0\;,
\eeq
on each end. 
To consider D-string, 
we impose the following conditions.
\beq
\delta \widetilde{X}^0, \delta \widetilde{X}^1 \neq 0, \delta \widetilde{X}^2 = \cdots = \delta \widetilde{X^9}=0\;.
\eeq
Therefore
\beq
\cases{
\partial_\sigma \widetilde{X}_0 -2 \partial_\tau \widetilde{A}_0^{0,\pi} =0 & \cr
\partial_\sigma \widetilde{X}_1 -2 \partial_\tau \widetilde{A}_1^{0,\pi} =0 & \cr
\partial_\tau \widetilde{X}^2=\cdots=\partial_\tau \widetilde{X}^9 =0 & \cr
}\;.
\eeq
We consider the following constant field strength of the gauge field.
\beq
F_{\alpha\beta}=
\pmatrix{0 & f_1 \cr -f_1 & 0 \cr} \;,\n
\widetilde{A}_\alpha^{0,\pi}=-\frac{1}{2} F_{\alpha\beta} \widetilde{X}^\beta\;,
\;\;\widetilde{A}_{\mu \neq \alpha}^{0,\pi}=0\;,
\eeq
where $F_{\alpha \beta}$ is the same in each brane 
since we are considering identical branes.\\
Now, the boundary condition is 
\beq
\cases{
\partial_\sigma \widetilde{X}_0 + f_1 \partial_\tau \widetilde{X}^1 =0 & \cr
\partial_\sigma \widetilde{X}_1 - f_1 \partial_\tau \widetilde{X}^0 =0 & \cr
\partial_\tau \widetilde{X}^2=\cdots=\partial_\tau \widetilde{X}^9 =0 & \cr
}\;,
\eeq
on each end. \\
We can rewrite the boundary condition as 
\beq
\left(
\begin{array}{cc}
\begin{array}{cc}
1 & 0 \\
0 & 1 
\end{array}
& {\bf 0} \\
{\bf 0} & {\bf 0}
\end{array} 
\right)
\partial_\sigma
\left(
\begin{array}{c}
\widetilde{X}_0 \\
\widetilde{X}_1 \\
\vdots
\end{array}
\right)
+ 
\left(
\begin{array}{cc}
\begin{array}{cc}
0 & f_1 \\
-f_1 & 0 
\end{array}
& {\bf 0} \\
{\bf 0} & {\bf 1}
\end{array} 
\right)
\partial_\tau 
\left(
\begin{array}{c}
\widetilde{X}^0 \\
\widetilde{X}^1 \\
\vdots 
\end{array}
\right)=0\;. \label{BC}
\eeq 
By using the fact that 
\beq
\widetilde{X}^1=\widetilde{X}_1,\;\widetilde{X}^0=-\widetilde{X}_0\;, 
\eeq
because of Minkowski space-time, we can rewrite eq. (\ref{BC}) again as
\footnote{
We sometimes omit indices.
Eq. (\ref{bcp=1}) means 
\beqn
\partial_{\sigma^+} \widetilde{X}_L^\mu (\sigma^+)\Big|_{\sigma=0,\pi}
={M_0}_{\;\nu}^\mu \partial_{\sigma^-} \widetilde{X}_R^\nu (\sigma^-)\Big|_{\sigma=0,\pi}\;.
\eeqn
}
\beq
\partial \widetilde{X}_L 
= M_0 \partial \widetilde{X}_R\;,\;\;
M_0
:=\pmatrix{
\matrix{ 
\frac{1+f_1^2}{1-f_1^2} & \frac{2f_1}{1-f_1^2} \cr 
\frac{2f_1}{1-f_1^2} & \frac{1+f_1^2}{1-f_1^2} 
} 
& {\bf 0} \cr {\bf 0} & {\bf -1}
}\;, \label{bcp=1}
\eeq
 where we decomposed the coordinate functions by using the equations of motion
\beq
\widetilde{X}(\tau, \sigma) = \widetilde{X}_L(\sigma^+) + \widetilde{X}_R(\sigma^-)\;,\;\;
\sigma^{\pm}:= \tau \pm \sigma \;.
\eeq
Then finally, we can get the following formula for the boundary condition in the original coordinate flame:
\beq
\cases{\partial X_L = M_v \partial X_R & $\cdots \sigma=0$ \cr
       \partial X_L = M_{-v} \partial X_R & $\cdots \sigma=\pi $}\;,\;\;
M_v:=\Lambda_v^{-1} M_0 \Lambda_v\;.
\eeq
\subsection{mode expansions for general boundary conditions}
We can treat general boundary condition with two D$p$-branes as follows.
\beq
\cases{
\partial X_L(\tau)=M \partial X_R(\tau) & $\cdots \sigma=0$ \cr
\partial X_L(\tau + \pi)=\bar{M} \partial X_R(\tau-\pi)  & $\cdots \sigma=\pi$ 
\cr
}\;,
\eeq
where $M$ and $\bar{M}$ are any $SO(D-1,1)$ matrices.
In this section, we proceed in this general case for a while, and sometimes come back to our special case: two identical parallel D$p$-branes' scattering.
Next, we have to  perform mode expansions of the coordinate functions
 which satisfy the above boundary condition.
The result is as follows.
\beq 
X(\tau,\sigma)=x + \sqrt{\alpha^\prime \over 2} \sum_{m \in {\bf Z}} 
\left[ M \int_0^{\sigma^+} + \int_0^{\sigma^-} \right] d\rho 
e^{-i\left( m+iE \right)\rho} \alpha_m \;,
\label{ModeEXP}
\eeq
where
\beq
E:= \frac{1}{2\pi} \ln{ M^{-1} \bar{M} } \;,
\eeq
is restricted to be diagonalizable.
$x$ is the integration constant for $\tau$ and $\sigma$,
which may depend on $\alpha_m$'s. 
\subsection{quantization}
We perform the canonical quantization by applying the canonical commutation relations to the coordinates and their conjugate momenta. 
The only non-zero commutation relations are, 
\beq
\left[ \partial_\tau X^\mu (\tau, \sigma),X^\nu (\tau, \sigma^\prime) \right]
= -2\pi i \alpha^\prime \delta(\sigma-\sigma^\prime)\eta^{\mu \nu},\;(0 < \sigma,\sigma^\prime \leq \pi) \;, \label{CCR}
\eeq
where 
\beq
\eta=\mbox{diag}(-1,1,\cdots,1)\;.
\eeq
We can get the commutation relations between modes from eq. (\ref{CCR}).\\
The results are
\footnote{See Appendix A.}
\footnote{For any Lorentz vectors $a^\mu$ and $b^\nu$, matrix $[a ,b]$ 
is defined by its elements to be $[a^\mu ,b^\nu]$.} 
\beq
\left[\alpha_m , \alpha_n \right]  &=& \left(m + i E \right) \eta \delta_{m+n} 
\;, \label{amam} \\ 
\left[x,\alpha_m \right]  &=& i\sqrt{\alpha^\prime \over 2}\left(1+M \right)\eta \;, \label{xam}\\
\left[x,x \right]  &=& \frac{\pi i \alpha^\prime}{2}\left( M-M^{-1}\right)\eta
\;\;. \label{xx}
\eeq
Note that the component of the matrix $M$ in the pure Neumann or pure Dirichlet direction are $1$ or $-1$, respectively. 
We can rewrite these relations as a more convenient form 
\footnote{See Appendix B. 
We would like to have complete pairs of canonical 
 variables and their conjugate momenta with one to one correspondence, 
by transforming 
$(x, \alpha_m) \rightarrow (\breve{x}, \breve{\alpha}_m)$.}
\footnote{
In this section and appendix B, 
we write any matrices A as: 
\beqn
A=\pmatrix{A_{11} & A_{12} \cr A_{21} & A_{22} \cr}\;,
\eeqn
where the dimension of the square matrix $A_{11}$ 
is defined to be equal to the rank of $E$.
}
\footnote{$t$ means transpose.}
.
\beq
\left[\breve{\alpha}_m , \breve{\alpha}_n \right] 
&=& \pmatrix{m+ iE_{11} & 0 \cr 0 & m \cr}\eta_T \delta_{m+n} 
\;, \label{baba} \\
\left[\breve{x},\breve{\alpha}_m \right]
&=&i\sqrt{\alpha^\prime \over 2}\pmatrix{0 & 0 \cr 0 & N_{22}}\eta_T\delta_m 
\;, \label{bxba} \\
\left[\breve{x},\breve{x} \right]
&=&\frac{\pi i \alpha^\prime}{2}
\pmatrix{
{
N_{11} \coth \pi E_{11} {\eta_{T}}_{11} N_{11}^t 
\atop 
+ N_{11} {\eta_{T}}_{11} (ST)_{11}^t - (ST)_{11} {\eta_{T}}_{11} N_{11}^t
}
 & 0 \cr 0 & 0\cr
} 
\;,\label{bxbx}
\eeq
where
\beq
\breve{\alpha}_m &:=& T^{-1}\alpha_m \;, \\
\breve{x} &:=& S x - \sqrt{\alpha^\prime \over 2} \sum_{m \in {\bf Z}} C_m \breve{\alpha}_m \;, \\
\eta_T &:=& T^{-1}\eta T^{-t} \;. \label{etat}
\eeq
In equations (\ref{baba})-(\ref{etat}), 
matrices $T$, $S$, $N$ and $C_m$ are defined by $M$ and $E$ in the appendix B. In the case of two identical parallel D$p$-brane scattering, 
the commutation relations are as follows
\beq
\left[\breve{\alpha}_m, \breve{\alpha}_n\right] 
&=& \pmatrix{\matrix{0 & m+i \epsilon \cr m-i \epsilon  & 0 \cr}  & {\bf 0} \cr {\bf 0} & m{\bf 1} \cr}\delta_{m+n} 
\;,\;\;
\epsilon := \frac{1}{\pi} \ln\left( g +v \over g -v \right) \;,
\n
\left[\breve{x}^i,p^j\right] 
&=& i\delta^{ij}\;,\;\;
\left[\breve{x},\breve{\alpha}_{m \neq 0}\right] 
=\big[\breve{x},\breve{\alpha}_0^{\mu \neq i}\big]
=\left[\breve{x},\breve{x}\right]
= 0 
\;,\n
\breve{\alpha}_0^i 
&=:& \sqrt{2\alpha^\prime} \mbox{diag}\Big( \frac{1}{g}, {\bf 1} + m_2^t, \cdots ,{\bf 1} + m_{p+1 \over 2}^t \Big)^{ij} p^j 
\;,\n
g &:=& \sqrt{1-f_1^2(1-v^2)} \;,\;i,j=2,\cdots, p+1 
\;,\label{our case}
\eeq
where $v$ is the velocity of the D-branes in the $(p+1)$-th direction 
and  $m_2,\cdots, m_{p+2 \over 2} $ are the $2\times2$ matrices 
in the following form.
\beq
m_1:=\pmatrix{\frac{1+f_1^2}{1-f_1^2} & \frac{2f_1}{1-f_1^2} \cr 
 \frac{2f_1}{1-f_1^2} & \frac{1+f_1^2}{1-f_1^2}}
\;,\;\;
m_l:=\pmatrix{\frac{1-f_l^2}{1+f_l^2} & \frac{-2f_l}{1+f_l^2} \cr  
\frac{2f_l}{1+f_l^2} & \frac{1-f_l^2}{1+f_l^2}}\;,\;\;l=2,\cdots,{p+1 \over 2}
\label{m_l}
\eeq
In eq. (\ref{m_l}) $f_1,\cdots,f_{p+1 \over 2}$ are the components of the U(1) gauge field strength:
\beq
F_{\alpha\beta}&:=&\bordermatrix{
& {\scriptstyle \beta=0 \;\;\;\;\;\;\;\;}  & \cdots & {\;\;\;\;\;\;\;\;\;\;\;\;\; \scriptstyle \beta=p} \cr
& \matrix{0 & f_1 \cr -f_1 & 0} & & \cr 
& & \ddots & \cr 
& & & \matrix{0 & f_{p+1 \over 2} \cr -f_{p+1 \over 2} & 0 \cr } \cr 
}_{\alpha\beta}\;, \label{F}
\eeq
defined in the proper flame of branes, as in the subsection 2.1 
\subsection{definition of the vacuum}
We consider two identical parallel D$p$-branes scattering here.
To define normal ordered operators, for instance, Virasoro operators,
we identify the creation and the annihilation operators with respect to the vacuum.  
Since $\epsilon \neq 0$, the commutation relations for the modes include the following unusual relation as compared with the case of pure Neumann or pure Dirichlet.
\beq
\left[\breve{\alpha}_0^0, \breve{\alpha}_0^1 \right]=i \epsilon .
\eeq
Therefore we classify the modes as
\beq
\cases{
\breve{\alpha}^\mu_{m>0} \;,\;\; \breve{\alpha}^1_0 & $\cdots$ annihilation operators \cr
\breve{\alpha}^\mu_{m < 0} \;,\;\; \breve{\alpha}^0_0 & $\cdots$ creation operators \cr
}
\;. \label{crea&anni}
\eeq
That is, the vacuum $|0>$ is defined as 
\beq
\breve{\alpha}^1_0 |0>= <0| \breve{\alpha}^0_0 = \breve{\alpha}^\mu_{m>0} |0>= <0|\breve{\alpha}^\mu_{m < 0} =0\;.
\eeq
Since exchange the role of $\breve{\alpha}^0_0$ and $\breve{\alpha}^1_0$ 
correspond to exchange the sign of the imaginary part of the Hamiltonian,
we choose the role as above in order to the imaginary part to be negative.
(Recall that $v$  is assumed to be positive.)
\subsection{Virasoro algebra}
We define the energy momentum tensor as 
\beq
T_{a b}:=-4\pi \alpha^\prime \frac{1}{\sqrt{h}}\frac{\delta S_{\rm bosonic}}{\delta h^{a b}}\;,
\eeq
where $S_{\rm bosonic}$ is the action in eq. (\ref{S_bosonic}) before the conformal gauge is chosen.
We introduce the holomorphic energy momentum tensor,
\beq
T(z):=\frac{1}{2 \alpha^\prime z^2} \left(T_{00} + T_{01} \right)\;,
\eeq
where
\beq
z:=e^{i \sigma^{\pm}}\;.
\eeq
If we define 
\beq
\widetilde{T(z)}:= T(z) - \frac{i\epsilon (1+i \epsilon)}{2 z^2}\;,
\eeq 
then
\begin{equation}
\widetilde{T(z)} \widetilde{T(w)} = \frac{\frac{D}{2}}{(z-w)^4} + \frac{2 \widetilde{T(w)} }{(z-w)^2}+ \frac{\partial \widetilde{T(w)}}{z-w} + \mbox{regular}
\;.
\footnote{See Appendix C.}
\end{equation}
This is the same form with the simpler boundary conditions. \\
Virasoro operators are defined by
\beq
\widetilde{L}_m:=\int_{|z|=1} \frac{d z}{2\pi i}z^{m+1} \widetilde{T(z)}\;.
\eeq
Now we can find the Virasoro algebra:
\beq
\left[\widetilde{L}_m , \widetilde{L}_n \right] &=& \int \frac{d w}{2\pi i}w^{n+1} \Res_{z=w} z^{m+1} \widetilde{T(z)} \widetilde{T(w)} \n
&=&(m-n)\widetilde{L}_{m+n} + \frac{D}{12} m(m^2-1) \delta_{m+n}\;.
\eeq
Note that the physical state condition by $L_0$ is different from the usual one, namely 
\beq
&& \left(\widetilde{L}_0 - c \right)\left| \mbox{phys}\right \rangle  \n
&=& \left( \frac{1}{2} \sum_{n \in {\bf Z} } :\alpha _{-n}^t \eta\alpha_n: 
- \frac{i \epsilon (1+ i \epsilon )}{2} - c \right)\left| \mbox{phys}\right 
\rangle \n
 &=& 0\;,
\eeq
where $:\;:$ means the normal ordering defined in the previous subsection and 
$c$ is the intercept, 
which is $1$ in the bosonic open string theory.
Now we can calculate the one loop scattering amplitude for bosonic string.
The formula is~\cite{BP} 
\beq
A_{\rm bosonic}
&=&-\ln{{\rm det}^{-\frac{1}{2}}\left(\widetilde{L}_0-c \right)}
= \frac{1}{2} \TTr{\ln{\left(\widetilde{L}_0-c \right) }} \n
&=& -\frac{1}{2} \int_0^{\infty} \frac{d t}{t} \TTr{e^{-\pi t \left(\widetilde{L}_0-c \right)}}\;. 
\eeq
However  we do not calculate this and we consider the superstring case in the next section. 
\section{D$p$-brane-D$p$-brane scattering in superstring theory}
\setcounter{equation}{0}
In this section we consider supersymmetric case.
Let us introduce super partners for dynamical variables $X$ in the previous section. We choose RNS formalism in this paper and introduce  fermion variable $\psi$ in the world sheet. In the super conformal gauge, 
the total action is
\footnote{
Our convention of two dimensional Dirac matrix is
\[
\rho^0=
\left(
\begin{array}{cc}
0 & -i \\ 
i & 0 
\end{array}
\right) \;,\;\; 
\rho^1=
\left(
\begin{array}{cc}
0 & i \\ 
i & 0 
\end{array}
\right)\;.
\]
}
\beq
S_{\rm string} &=& -\frac{1}{4 \pi \alpha^\prime} \int d\sigma d\tau \left[ \partial_a X^\mu \partial^a X_\mu  -i \bar{\psi}^\mu \rho^a \partial_a \psi_\mu \right]
\n
&+& \frac{1}{4 \pi \alpha^\prime} \left.\int d\tau F^v_{\mu\nu}
\left[ X^\nu \partial_\tau X^\mu - \frac{i}{2} \bar{\psi}^\nu \rho^0 \psi^\mu \right]\right|_{\sigma=0}^{\sigma=\pi} \n
&+& ({\rm ghost}) \;, 
\label{Sstring}
\eeq
where
\beq
F^v_{\mu \nu}&=&{\Lambda_{\pm v}}^\rho_{\;\mu} {\Lambda_{\pm v}}^\sigma_{\;\nu}
\pmatrix{F & 0 \cr 0 & 0 \cr}_{\rho\sigma}
\;, \label{F^v}
\eeq
where $\pm$ correspond to $\sigma=0$ and $\sigma=\pi$, respectively and 
\beq
{\Lambda_v}^\rho_{\;\mu}=\bordermatrix{
& {\scriptstyle \mu=0} & {\scriptstyle \cdots} & {\scriptstyle \mu = p+1} & {\scriptstyle \cdots} \cr 
& \frac{1}{\sqrt{1-v^2}} &  {\bf 0}  & \frac{-v}{\sqrt{1-v^2}} & \cr
& {\bf 0} & {\bf 1}  & {\bf 0}  & {\bf 0} \cr
& \frac{-v}{\sqrt{1-v^2}}   & {\bf 0}  & \frac{1}{\sqrt{1-v^2}} & \cr
& & {\bf 0} & & {\bf 1} \cr
}^\rho_{\;\;\;\mu}\;.
\eeq
In eq. (\ref{F^v}), $F$ defined in eq. (\ref{F}) is a field strength on D-branes world volume and is set to be constant.
Since we are considering identical branes, $F$ is the same in each brane.
$\Lambda_v$ is a Lorentz boost matrix 
with the velocity $v$ in the ($p+1$)-th direction. \\
Since the spinor is written as follows
\beq
\psi= \left( \psi_R \atop \psi_L \right)\;, 
\eeq
the world sheet global supersymmetry is expressed as
\beq
\cases{
\delta_{\rm susy} X^\mu = i \epsilon_{\rm susy} \left( \psi_L^a  + \psi_R^a \right)^\mu &\cr
\delta_{\rm susy} \psi _{L \atop R}^{a \mu} = -2 \epsilon_{\rm susy} e^{i (1-a) \sigma_{\pm}} \partial_{\pm} X^\mu
}\;,
\eeq
where $\epsilon_{\rm susy}$ is independent of $\tau$,$\sigma$ and
\beq
\psi_{L \atop R}^{a} := e^{i \frac{1-a}{2} \sigma_\pm} \psi_{L \atop R}
\;\;\;\cdots
\cases{
a=0 & NS sector \cr
a=1 & R sector \cr
}\;.
\eeq
Due to the existence of the boundary (i.e. D-brane), 
there is only one parameter for supersymmetry.
\subsection{boundary condition}
As in bosonic case, in order to get free equations of motion, 
we require
\beq
\delta X \left( \eta \partial^1 X - F^v \partial_0 X  \right) + \frac{i}{2} \delta \bar{\psi} \left( \eta \rho^1 \psi + F^v \rho^0 \psi \right) =0\;,
\eeq
or represent equivalently as
\beq
\delta X \left[ \left( \eta +F^v \right) \partial X_L
-\left( \eta -F^v \right)\partial X_R \right]&=&0   \label{bos} \;, \\
\delta \psi_L\left( \eta +F^v \right) \psi_L 
- \delta \psi_R \left( \eta -F^v \right)\psi_R&=&0\;. \label{fer}
\eeq
In the proper time coordinate frame of D-branes, 
we impose the  following boundary conditions 
for bosons and fermions at the both ends.
\footnote{
We determine boundary conditions for fermions as follows.
If the boundary condition for bosons is
\beqn
\left( \partial_+  - M \partial_- \right) X=0\;,
\eeqn
from eq. (\ref{bos}) as in the section 2, 
then we transform this by the supersymmetry and get the following equation. 
\beqn
\left( \partial_+  - M \partial_- \right) \left( \psi_L^a + \psi_R^a \right)=0\;.
\eeqn
By using the equations of motion for fermions: $\partial_{\pm}\psi_{R \atop L}^a=0$, 
we get 
$
\partial_{\tau}\left( \psi_L^a - M \psi_R^a \right)=0
$. 
Then we impose the following boundary condition for fermions. 
\beqn
\psi_L^a = M \psi_R^a\;.
\eeqn
Eq. (\ref{fer}) determines the  relation between $\delta \psi_L$ and  $\delta \psi_R$.
}
\beq
\cases{
\begin{array}{lcl}
(\eta + F) \partial \widetilde{X_L} &=& (\eta -F) \partial \widetilde{X_R}  \cr
(\eta + F)\widetilde{\psi_L^a} &=& (\eta -F)\widetilde{\psi_R^a} \cr
\end{array}
  & $\cdots$ Neumann direction \cr
\begin{array}{lcl}
\partial \widetilde{X_L} &=& -\partial \widetilde{X_R}  \cr
\widetilde{\psi_L^a} &=& -\widetilde{\psi_R^a}  \cr
\end{array} 
& $\cdots$ Dirichlet direction \cr
}\;.
\eeq
In the original frame, that is
\beq
\cases{
\begin{array}{l} 
\partial X_L=M_v \partial X_R \\
\psi_L=M_v \psi_R 
\end{array} & $(\sigma=0)$ \cr
\begin{array}{l}
\partial X_L =M_{-v} \partial X_R \\
\psi_L =(-)^{1-a}M_{-v}\psi_R
\end{array} & $(\sigma=\pi)$ \cr
} \label{bcsusy} \;.
\eeq
In eq. (\ref{bcsusy}), 
\beq
M_v:=\Lambda_v^{-1} M_0 \Lambda_{v}\;,
\eeq
\beq
M_0:=\pmatrix{m_1 &&&&{\bf 0} \cr& \ddots &&  &\cr & && m_{p+1 \over 2 } &\cr{\bf 0} &&&& {\bf -1} \cr}\;,
\eeq
where $m_1 ,\cdots, m_{p+1 \over 2}$ are $2\times 2$ matrices which are defined in the previous section.
\subsection{mode expansion}
We can find the mode expansions for $X$ and $\psi$ to satisfy the above boundary condition as follows.
\beq
X(\tau,\sigma)&=&x + \sqrt{\alpha^\prime \over 2} \sum_{m \in {\bf Z}} 
\left[ M_v \int_0^{\sigma^+} + \int_0^{\sigma^-} \right] d\rho 
e^{-i\left( m+iE \right)\rho} \alpha_m \;,
\\
\psi_L(\sigma^+)&=&
\sqrt{\alpha^\prime} M_v \sum_{r\in {\bf Z}-\frac{1-a}{2}} e^{-i \left(r + i E \right)\sigma^+} d_r \;,
\\
\psi_R(\sigma^+)&=&
\sqrt{\alpha^\prime} \sum_{r\in {\bf Z}-\frac{1-a}{2}} e^{-i \left(r + i E \right)\sigma^-} d_r \;,
\eeq
where 
\beq
E:= \frac{1}{2\pi} \ln{ M^{-1}_v M_{-v} }  \;,
\eeq
which is diagonalizable.
\subsection{quantization}
We can find that the commutation relations for fermionic modes $d$ are the usual type:
\beq
\left\{ d_r, d_s \right\} = \eta \delta_{r+s}\;,
\eeq
and bosonic commutators is as in the previous section:
\beq
\left.\matrix{
\left[\alpha_m , \alpha_n \right] & = & \left(m + i E \right) \eta \delta_{m+n}   \cr
\left[x,\alpha_m \right] & = & i\sqrt{\alpha^\prime \over 2}\left(1+M_v \right)\eta   \cr
\left[x,x \right] & = & \frac{\pi i \alpha^\prime}{2}\left( M_v-M_v^{-1}\right)\eta  \cr
}\;\;.
\right. \label{com2}
\eeq
Eq. (\ref{com2}) is rewritten as eq. (\ref{our case}).
The ghost part is the same as usual because the boundary condition does not change it.
\subsection{Virasoro algebra}
The energy momentum tensor is
\beq
T(z) &:=& \frac{1}{\alpha^\prime z^2} 
\left[
\partial_- X^\mu \partial_- X_\mu + \frac{i}{2} \psi_R^\mu \partial _- \psi_{R \mu} 
\right] + ({\rm ghost}) \n
&=&T_{\rm boson}(z)+ T_{\rm fermion}(z) + T_{\rm ghost}(z) \n
&:=& \frac{1}{2} \sum_{m,n \in {\bf Z}} z^{-m-n-2} :\alpha_m^T \eta \alpha_n:\n
&+&\frac{1}{2} \sum_{r,s \in {\bf Z} -\frac{1-a}{2}} z^{-r-s-2} :d_r^T \eta (s+iE) d_s: \n
&+& ({\rm ghost})\;.
\eeq
If we define a shifted energy momentum tensor as
\beq
\widetilde{T(z)}&:=& T(z) - \frac{i\epsilon (1+i \epsilon)}{2 z^2} + \frac{aD - 8\epsilon^2}{16 z^2} \n
&=&T(z) + \frac{1}{z^2}\left( \frac{a D}{16} - \frac{i \epsilon}{2} \right) \;,
\label{EMT}
\eeq
then we get
\footnote{See appendix C.}
\beq
\widetilde{T(z)} \widetilde{T(w)} 
&=& \frac{\frac{3}{4} D}{(z-w)^4} + \frac{2 \widetilde{T(w)}}{(z-w)^2} + \frac{\partial_w \widetilde{T(w)}}{z-w} \n
&+& ({\rm regular} + {\rm ghost \;\; contribution})\;.
\label{TT}
\eeq
This form is of usual type. 
\subsubsection{physical state condition}
Virasoro operators are defined by
\beq
L_m:=\int_{|z|=1} \frac{d z}{2\pi i}z^{m+1} T(z)\;.
\eeq 
Note that the physical state condition by $L_0$ is different from the usual one
because energy momentum tensor $\widetilde{T}$ which satisfies the correct operator product expansions is shifted from $T$ as eq. (\ref{EMT}).
That is,
\beq
\left(L_0 -\frac{i \epsilon}{2} -\frac{1-a}{2} \right)\left| \mbox{phys}\right \rangle =0\;.
\eeq
We further define 
\beq
L_0^a&:=&L_0 -\frac{i \epsilon}{2} -\frac{1-a}{2} \n
&=& L_0^{\alpha} + L_0^{\rm d} + L_0^{\rm bc} + L_0^{\beta \gamma} -\frac{i \epsilon}{2} +\frac{1-a}{2}\;,
\eeq
\beq
L_0^{\alpha}&:=&\frac{1}{2} \sum_{n \in {\bf Z} } :\alpha _{-n} \eta \alpha_n: 
\;, \label{L0a} \\
L_0^{\rm d}&:=&\sum_{r=1-\frac{1-a}{2}}^{\infty} d_{-r}^t \eta (r+iE) d_r + \frac{a}{2} d_0^t \eta i E d_0 \;,\\ 
L_0^{\rm b c}&:=&\sum_{n=1}^{\infty} n\left( b_{-n} c_n + c_{-n} b_n  \right) 
\;,\\
L_0^{\beta \gamma}&:=&\sum_{r=1-\frac{1-a}{2}}^{\infty} r \left( \beta_{-r} \gamma_r - \gamma_{-r} \beta_r  \right) \;,
\eeq
where $b$, $c$, $\beta$, and $\gamma$ are the standard ghosts.
The normal ordering procedure in eq. (\ref{L0a}), 
is found by eq. (\ref{crea&anni}).
\subsection{scattering amplitude}
We calculate the scattering amplitude for two D$p$-branes, 
namely, 
the one loop amplitude of open string 
whose ends are attached on the D$p$-branes.
The one loop amplitude can be expressed by the following proper time ($t$) integral of the exponentiated Hamiltonian, 
which annihilate the physical state. 
It is gotten in the previous subsection, 
that is $L_0$ with the correct constant shift. 
The shift can be found by imposing the physical state condition 
in the previous subsection. 
Therefore we use $L_0^a$ as Hamiltonian. 
The integral region is from $0$ to $\infty$ because of cylinder~\cite{ACNY}. 
Since we are choosing the NSR formalism, 
we have to sum over the spin structures which have $2 \times 2=4$ combinations because of the cylinder amplitude. 
The formula of the amplitude is~\cite{BP}
\beq
A_{\rm string} &=& -\ln{{\rm det}^{-\frac{1}{2}}L_0^a}= \frac{1}{2} \TTr{\ln{L_0^a}}=-\frac{1}{2} \int_0^{\infty} \frac{d t}{t} \TTr{e^{-\pi t L_0^a}} \n
&=& -\frac{1}{2} \int_0^{\infty} \frac{d t}{t} \sum_{a,b=0,1} C\left(a \atop b \right) \Tr_{\rm NSR}{(-)^{b F} e^{- \pi t L_0^a}}\;, \label{astring0}
\eeq
where $F=F^{\rm d}+ F^{\rm \beta,\gamma}$ is the fermion number operator, 
the summation is done over the spin structures which consist of GSO projection ($b=0,1$)  
and  the summation of NS ($a=0$) and R ($a=1$) sectors.
 $C{a \choose b}$ are their weights:
\beq
C{0 \choose 0}=-C{0 \choose 1} =-C{1 \choose 0}=\frac{1}{2} 
\;.
\eeq
"${\displaystyle \Tr_{\rm NSR}}$" means 
\beq
\Tr_{\rm NSR} = 2 \times \Tr_\alpha \times \Tr_{\rm d} \times \Tr_{\rm b,c} \times \Tr_{\beta,\gamma}\;,
\eeq
where the factor $2$ is caused by the sum of the configurations of the exchanged ends of open string. 
Therefore
\beq
\Tr_{\rm NSR}{(-)^{b F} q^{L_0^a}} &=& 2 \times q^{-\left(\frac{i \epsilon}{2} + \frac{1-a}{2}\right)} \n
&& \times \Tr_{\alpha}q^{L_0^\alpha} 
   \times \Tr_{\rm b,c}q^{L_0^{\rm b,c}} \n
&& \times \Tr_{\rm d}(-)^{b F^{\rm d}}q^{L_0^{\rm d}}
   \times \Tr_{\rm \beta, \gamma}(-)^{b F^{\rm \beta,\gamma}}q^{L_0^{\rm \beta,\gamma}} \;,
\eeq
where 
\beq
q:=e^{-\pi t}\;.
\eeq
We calculate these traces as follows.
\beq
\Tr_{\alpha}q^{L_0^{\rm \alpha}} 
&=& q^{{1 \over 2} \sum_{k=p+2}^{D-1} \left.\breve{\alpha}_0^k \right.^2}
\times \Tr_{\rm p^i}{q^{{1 \over 2} \sum_{i=2}^{p+1} \left.\breve{\alpha}_0^i \right.^2}}
\times \Tr_{\breve{\alpha}_0^{1,0},\breve{\alpha}_{n \neq 0}}{q^{L_0^{\rm \alpha} \big|_{\mbox{ \scriptsize frequency term}} }} \n
&=& q^{b^2 \over 4\pi^2 \alpha^\prime}
\times V_p \int_{-\infty}^{\infty} \prod_{i=2}^{p+1} \frac{dp^i}{2\pi} q^{{1\over 2} \left.\breve{\alpha}_0^i \right.^2} \n
&\times& \left(1-q^{-i\epsilon}\right)^{-1} \prod_{n=1}^{\infty} \left(1-q^n \right)^{-(D-2)} \left| 1-q^{n-i\epsilon}\right|^{-2} 
\;,\\
\Tr_{\rm b,c}q^{ L_0^{\rm b,c}} 
&=& \prod_{n=1}^{\infty} \left( 1- q^n \right)^2 
\;,\\
\Tr_{\rm d}(-)^{b F^{\rm d}}q^{L_0^{\rm d}} 
&=& (1-a b)\left( 2^4 \cos{\pi t \epsilon \over 2}\right)^a \n
&\times& \prod_{n=1}^{\infty} \left( 1+ (-)^b q^{n-\frac{1-a}{2}}\right)^{(D-2)} \left| 1+(-)^b q^{n-\frac{1-a}{2} -i \epsilon}\right|^2 
\;,\\
\Tr_{\rm \beta, \gamma}(-)^{b F^{\rm \beta,\gamma}}q^{ L_0^{\rm \beta,\gamma}} 
&=& \prod_{n=1}^{\infty} \left( 1+ (-)^b q^{n-\frac{1-a}{2}}\right)^{-2} \;.
\eeq
In the above equations, the zero mode integral 
\footnote{$V_p$ is the space volume of a D$p$-brane. See appendix B and see also Ref.~\cite{ACNY}.}
can be done by using the relation between $p^i$ and $\breve{\alpha}^i$ obtained in the previous section.
The result can be expressed as follows by using Jacobi's theta functions: $\theta\left( a \atop b \right)$ 
\footnote{The definition and some properties are 
in an appendix in Ref.~\cite{K}.},
\beq
\Tr_{\rm \alpha,b,c}q^{\left( L_0^\alpha + L_0^{\rm b,c} \right)} 
&=&V_p g \prod_{l=2}^{p+1 \over 2}(1+f_l^2)\left(4\pi^2 \alpha^\prime t\right)^{-\frac{p}{2}} e^{-\frac{b^2}{4 \pi \alpha^\prime} t} \n
&\times& \frac{i q^{ i \epsilon \over 2} q^{1 \over 8}}{\prod_{n=1}^{\infty} \left( 1-q^n\right)^5}
\times \frac{1}{\theta\left(1 \atop 1 \right) \left( \left. \epsilon t \over 2 \right.\left| i t \over 2 \right. \right)} 
\;,\\
\Tr_{\rm d,\beta,\gamma}(-)^{b F}q^{ \left( L_0^{\rm d} + L_0^{\beta,\gamma}\right)} &=& \frac{\left.\theta\left( a \atop b \right) \left( \left. 0 \right.\left| i t \over 2 \right. \right) \right. ^3 \theta\left( a \atop b \right) \left( \left.    \epsilon t \over 2 \right.\left| i t \over 2 \right. \right)}{q^{a \over 2} \prod_{n=1}^{\infty} \left( 1-q^n\right)^4} 
\;.
\eeq
Therefore we can find the scattering amplitude eq. (\ref{astring0}) as
\beq
A_{\rm string}
&=&-i(2\pi)^3 V_p  g \prod_{l=2}^{p+1 \over 2}(1+f_l^2) 
\cdot \int_0^{\infty} \frac{d t}{t} (4\pi^2 \alpha^\prime t )^{-\frac{p}{2}} e^{- \frac{b^2}{4\pi \alpha^\prime} t} \n
& \times & \sum_{a,b = 0,1} C\left(a \atop b \right)
\frac{\left.\theta\left( a \atop b \right) \left( \left. 0 \right.\left| i t \over 2 \right. \right) \right. ^3 
\theta\left( a \atop b \right) \left( \left.    \epsilon t \over 2 \right.\left| i t \over 2 \right. \right)}
{\left.\theta\left( 1\atop 1 \right)^\prime \left( \left. 0 \right.\left| i t \over 2 \right. \right) \right. ^3
\theta\left( 1 \atop 1 \right) \left( \left.    \epsilon t \over 2 \right.\left| i t \over 2 \right. \right)} \label{astring1} \;.
\eeq
Using Jacobi identity:
\beq
\sum_{a,b = 0,1} C\left(a \atop b \right)
\left.
 \theta\Big( {a \atop b} \Big) \Big( 0 \Big| {i t \over 2} \Big) 
\right. ^3 
\theta\Big( {a \atop b} \Big) \Big( {\epsilon t \over 2} \Big| {i t \over 2}\Big)
=
\left.
 \theta\Big( {1 \atop 1} \Big) \Big( {\epsilon t \over 4} \Big| {i t \over 2} \Big)\right. ^4 \;,
\eeq
and the property of the theta function under the modular transformation:
\beq
\matrix{
\theta\left( a \atop b \right) \left( \left.    \epsilon t \over 2 \right.\left| i t \over 2 \right. \right) 
&=& \sqrt{2 \over t} e^{\frac{i\pi a b}{2} + \frac{\pi t \epsilon^2}{2}} 
\theta\left( b \atop -a \right) \left(\left. i \epsilon \right.\left| 2i \over t \right.\right) \cr
\theta\left( 1 \atop 1 \right)^\prime \left(\left. 0 \right.\left| it \over 2 \right. \right)
&=& \left(2 \over t\right)^{3 \over 2} 
\theta\left( 1 \atop 1 \right)^\prime \left(\left. 0 \right.\left| 2 i \over t \right.\right) \cr
}
\;,
\eeq
we can further simplify the amplitude eq. (\ref{astring1}) as
\beq
A_{\rm string} 
&=&
 V_p g \prod_{l=2}^{p+1 \over 2}(1+f_l^2) 
\cdot \int_0^\infty \frac{d t}{t} (4\pi^2 \alpha^\prime t)^{-\frac{p}{2}} e^{-\frac{b^2}{4\pi \alpha^\prime}t} \n
&\times&
\frac{(\pi t)^3 }{\theta\left( 1\atop 1 \right)^\prime \left( \left. 0 \right.\left| 2i \over t \right. \right)^3}
\frac{\theta\left( 1\atop 1 \right) \left( \left. i \epsilon \over 2 \right.\left| 2i \over t \right. \right)^4}{ \theta\left( 1\atop 1 \right) \left( \left. i \epsilon \right.\left| 2i \over t \right. \right)}
\;. \label{astring2}
\eeq
\vfill
\section{D$p$-brane-D$p$-brane scattering in IIB matrix model}
\setcounter{equation}{0}
In this section, with our notation, we briefly review the calculation of the brane scattering amplitude in the IIB matrix model~\cite{M}. 
In this model, dynamical variables are "coordinates" represented by $N \times N$ matrices and $N$ is assumed to be infinitely large.\\
The action is
\beq
S_{\rm matrix}=
\alpha \Tr{\left( -\frac{1}{4}\left[ X_\mu, X_\nu \right]^2 
-\frac{1}{2} \bar{\psi} \Gamma^\mu \left[X_\mu,\psi \right]  \right)}
+ \beta N\;.
\eeq
The classical equations of motions for $X^\mu$ from this action are 
\beq
\left[X^\mu,\left[X_\mu,X_\nu \right]\right]=0\;,
\eeq
which has the solutions that may be representing D-branes.
In our situation: 
two identical parallel $p$-branes moving with relative velocity $2v$, 
the following classical configuration is suitable~\cite{M}.
\beq
X^\mu = \pmatrix{X^\mu_{(1)} & 0 \cr 0 & X^\mu_{(2)} \cr} \;,\;\; \psi=0\;,
\eeq  
where $X_{(I)}$ are ${N \over 2}\times {N \over 2}$ matrices 
and their indices $I=1,2$ mean the first and second branes.\\
First, in the proper time frame (written as $\tilde{X}$) for each brane, 
we set 
\beq
\left[ \tilde{X}^\alpha_{(I)},\tilde{X}^\beta_{(I)} \right]
=2 \pi i \alpha^\prime F^{M\;\alpha \beta} {\bf 1}_{{N \over 2}\times{N \over 2}} \;,\;\;\mbox{all other commutator}=0\;, \label{stcom} \\
F^M=\pmatrix{\matrix{0 & f_1^M \cr -f_1^M & 0} & & \cr & \ddots & \cr & & \matrix{0 & f^M_{p+1 \over 2} \cr -f^M_{p+1 \over 2} & 0}}\;,\;\;\det{F^M}\neq 0\;,
\eeq
in order to describe $p$-branes.\\
Since we are considering identical branes, $F^M$ is independent of $I=1,2$.
The right hand side of eq. (\ref{stcom}) 
is proportional to the identical matrix.
To satisfy the eq. (\ref{stcom}), 
we use $\frac{p+1}{2}$ pairs of matrices 
($q_l$ , $p_l$) with large enough dimensions.
\beq
\cases{
\tilde{X}^{2l-2}_{(I)}=\sqrt{  2 \pi \alpha^\prime f^M_{l}} q_l & \cr
\tilde{X}^{2l-1}_{(I)}=\sqrt{ 2 \pi \alpha^\prime f^M_{l}} p_l & \cr
\tilde{X}^{p+1}_{(I)} = \cdots =\tilde{X}^{D-2}_{(I)}=0 & \cr
\tilde{X}^{D-1}_{(I)}=\pm \frac{b}{2}
} 
\;,\;\;[q_l,p_{l^\prime}]= i \delta_{l l^\prime}
\;,\;\;l,l^\prime=1, \cdots, \frac{p+1}{2}
\;.
\eeq
Next, for the first and second branes, 
we perform the boost with velocity $v$ and $-v$
along the ($p+1$)-th axis, respectively as
\beq
\pmatrix{X^0_{(I)} \cr X^{p+1}_{(I)}}
=\pmatrix{
1 \over \sqrt{1-v^2} & \pm v \over \sqrt{1-v^2} \cr
\pm v \over \sqrt{1-v^2} & 1 \over \sqrt{1-v^2} \cr
}
\pmatrix{\tilde{X}^0_{(I)} \cr \tilde{X}^{p+1}_{(I)}}\;,
\eeq
where $\pm$ correspond to $I=1,2$.
We can get the following solution:
\beq
\cases{
X^0 = \sqrt{ 2 \pi \alpha^\prime f^M_1} \pmatrix{q_0 \frac{1}{\sqrt{1-v^2}} & 0 \cr 0 &  q_0 \frac{1}{\sqrt{1-v^2}}  \cr}
& \cr
X^{2l-2} = \sqrt{ 2 \pi \alpha^\prime f^M_l } \pmatrix{q_l  & 0 \cr 0 &  q_l   
\cr}
& $l=2, \cdots, \frac{p+1}{2}$ \cr
X^{2l-1} = \sqrt{ 2 \pi \alpha^\prime f^M_l } \pmatrix{p_l & 0 \cr 0 &  p_l  
\cr}
&   $l=1, \cdots, \frac{p+1}{2}$ \cr
X^{p+1}=\sqrt{ 2 \pi \alpha^\prime f^M_1} \pmatrix{q_0 \frac{v}{\sqrt{1-v^2}} & 0 \cr 0 & q_0 \frac{-v}{\sqrt{1-v^2}}  \cr}
& \cr
X^{p+2 \cdots D-2}=0 &  \Bigg.\Bigg. \cr
X^{D-1}= \pmatrix{b \over 2 & 0 \cr 0 & -\frac{b}{2}}
& \cr
}\;,
\eeq
where $2v$ is the relative velocity between the branes.\\
The result of the calculation for the effective action (scattering amplitude)
in the one loop level is~\cite{M}
\beq
A_{\rm matrix}
&=&
-i 8 V_p \prod_{l=2}^{p+1 \over 2} {f^M_l}^{-2} 
\cdot \int_0^\infty \frac{d t}{t} (4\pi^2 \alpha^\prime t)^{-\frac{p}{2}} e^{-\frac{b^2}{4\pi \alpha^\prime}t} \n
&\times&
\frac{1}{f_1^M \cosh{\epsilon^M}}
\frac{\sin^4{\left({f_1^M \sinh{\epsilon^M} \over 2}t \right)}}{\sin{\left(f_1^M \sinh{\epsilon^M }t\right) }}
\;,\;\;\; \label{amatrix}
\eeq
where 
\beq
\tanh{\epsilon^M}:=v\;,
\eeq
and the fact that the matrix size $N$ is large enough is used.
\vfill
\setcounter{equation}{0}
\section{Comparing string theory and matrix model}
We compare  the two scattering amplitudes in the IIB string theory and the IIB matrix model, respectively.
They can be rewritten as follows:
\beq
A_{\rm string}
&=&
\frac{-i}{2}V_p
\cdot \frac{\displaystyle \prod_{l=2}^{p+1 \over 2 } \left(f_l^2 +1\right)}{1-f_1^2}
\cdot {v^3 \over 1-v^2}
\cdot \int_0^\infty \frac{d t}{t} (4\pi^2 \alpha^\prime t)^{-\frac{p}{2}} e^{-\frac{b^2}{4\pi \alpha^\prime}t} \n
&\times &
\prod_{n=1}^{\infty}
\frac{\left[1-\left(\sinh{\pi \epsilon \over 2} \over \sinh{2 \pi n \over t}\right)^2 \right]^4}{1-\left(\sinh{\pi \epsilon} \over \sinh{2 \pi n \over t}\right)^2}
\;,\;\;
\sinh{\pi \epsilon}=\frac{2 v \sqrt{1-f_1^2(1-v^2)}}{(1-v^2)(1-f_1^2)}
\;.\;\;\; \label{string}
\eeq
\beqn
\Updownarrow  
\eeqn
\beq
A_{\rm matrix}
&=&
\frac{-i}{2}V_p
\cdot 
\frac{{f^M_1}^2}{ \displaystyle \prod_{l=2}^{p+1 \over 2} \left.f^M_l \right.^2}
\cdot {v^3 \over 1-v^2}
\cdot \int_0^\infty \frac{d t}{t} (4\pi^2 \alpha^\prime t)^{-\frac{p}{2}} e^{-\frac{b^2}{4\pi \alpha^\prime}t} \n
&\times &
\prod_{n=1}^{\infty}
\frac{\left[1-\left({f_1^M \sinh{\epsilon^M}} \over {2 \pi n \over t}\right)^2 \right]^4}{1-\left({2f_1^M \sinh{\epsilon^M}} \over {2 \pi n \over t}\right)^2}
\;,\;\;
\sinh{\epsilon^M}=\frac{v}{\sqrt{1-v^2}}
\;.\;\;\;\; \label{matrix}
\eeq
Here we compare the imaginary parts of the phase shifts; ${\rm Im}(-i A)$  as:
\beq
{\rm Im}(-i A_{\rm string})
&=&
8 V_p \sum_{k=1,3,5,\cdots} \frac{1}{k}
\left(\frac{{\epsilon \over 2}}{4 \pi^2 \alpha^\prime k} \right)^{p \over 2} 
e^{-\frac{b^2}{4\pi\alpha^\prime}\frac{2k}{\epsilon}} \n
 & \times &
\prod_{l=2}^{p+1 \over 2}\left(1+f_l^2\right) 
\cdot g \prod_{n=1}^{\infty}
\coth^8{\left(\pi k n \over \epsilon \right)}
\;.\;\;\;\label{IMS}
\eeq
\beqn
\Updownarrow  
\eeqn
\beq
{\rm Im}(-i A_{\rm matrix})
& = &
8 V_p \sum_{k=1,3,5,\cdots} {1 \over k} 
\left( 
 \frac{{f_1^M \sinh{\epsilon^M} \over \pi}}{4 \pi^2 \alpha^\prime k}
\right)^{p \over 2} 
e^{-{b^2 \over 4\pi \alpha^\prime}{\pi k \over f_1^M \sinh{\epsilon^M}}} \n
&\times&
\prod_{l=2}^{p+1 \over 2} \left. f_l^M \right.^{-2} 
\cdot \frac{\sqrt{1-v^2}}{f_1^M}
\;.\;\;\;\label{IMM}
\eeq
\subsection{three special limits}
Now, if we assume a relation between the U(1) field strength $F$ in string theory and the value of commutator $F^M$ among coordinates in IIB matrix model as follows:
\beq
\frac{\displaystyle \prod_{l=2}^{p+1 \over 2 } \left(f_l^2 +1\right)}{1-f_1^2}\cdot \frac{ \displaystyle \prod_{l=2}^{p+1 \over 2} \left.f^M_l \right.^2}{{f^M_1}^2}
=1 \;,\;\;\;\label{T-dual}
\eeq
considering three different regions of parameters
which are $b$, $v$ and $f_1$ as:
\begin{equation}
\cases{
b \gg \alpha^\prime & (long distance) \cr
v \ll 1 & (low velocity) \cr
\left|f_1 \right| \gg 1 &  (large field strength)
\cr
}
\footnote{
We put $f_1$ pure imaginary 
which has the large absolute value, 
that is small real $f_1^M$
}, 
\end{equation}
then, in each region, 
the both amplitudes approach the same value as follows: 
\beq
A_{\rm {string \atop matrix}}
\longrightarrow
\frac{-i}{2} \Gamma\left(3-{p \over 2}\right)
\cdot \frac{V_p}{\left(4\pi^2\alpha^\prime\right)^{p \over 2}}
\cdot \frac{\displaystyle \prod_{l=2}^{p+1 \over 2 } \left(f_l^2 +1\right)}{1-f_1^2}
\cdot {v^3 \over 1-v^2}
\cdot \left(\frac{4\pi\alpha^\prime}{b^2}\right)^{3-{p \over 2}}\!\!\!\;.
\eeq
In addition, 
in the case of $v \ll 1$ or $\left|f_1 \right| \gg 1$ 
the imaginary parts of the phase shifts; 
eq. (\ref{IMS}) and eq. (\ref{IMM}), 
are also identical.
%
\vspace*{5pt}\\
The main conclusion of this paper is that eq. (\ref{string}) and eq. (\ref{matrix}) are identical in three independent limit of parameters
after identifying the field strengths in the models as eq (\ref{T-dual}).
This condition is satisfied when we require
\beq
\det{(\eta +F)F^{M}}=-1\;.
\eeq
\vfill
\section{Conclusions and Discussions}  
\setcounter{equation}{0}
In this paper, 
we have considered two identical $p$-branes 
and compared their scattering amplitudes
in the IIB string theory and the IIB matrix model.
Here the $p$-brane has the lower brane charges in the various ways.
That is, in the context of the string theory, 
there are arbitrary constant electromagnetic field strength $F$ on the $p$-brane world volume.
On the other hand, in the context of the matrix model, 
there are various values ($F^M$) of the commutators among brane coordinates. 
Based on  the matrix model, 
we should calculate the amplitude 
around the classical back-ground of the two body system of $p$-branes.
However we have only the amplitude in the one loop level.
The loop expansion is assumed to be the low energy expansion.        
Therefore the corresponding value based on the string theory is a cylinder diagram in the low energy limit, 
which is attached on the nontrivial background.
It means the one loop calculation of open string 
with the corresponding boundary condition. 
After we have performed the low energy expansion 
that is the long distance expansion 
or small velocity  expansion in each model, 
or small $f_1^M$ expansion in IIB matrix model 
and large $|f_1|$ expansion in IIB string theory.
the leading terms of the amplitudes are found to be the same precisely
if we identify $\det{(\eta +F)F^{M}}=-1$.
When all $f^M_l$s are the same and $f_l=0$ in the all world volume directions, 
our result reproduces that in Ref.~\cite{M}.
When $f_{l}^M \ll 1$, 
our result reproduces that in Ref~\cite{CT,I}.
Note that we have performed 
these expansions of three parameters independently.
In this paper,
 we have obtained a  clear correspondence between the IIB string theory and its IIB matrix model in the near BPS configuration: D-brane scattering.
Since we also find a way to study general boundary conditions
we can calculate various configurations of two body systems of branes.
For example, D$p$-brane and D$q$-brane case for $p\neq q$ with different lower dimensional brane charges and with arbitrary angles.
We can also consider the relation between IIA string theory and matrix model for M-theory~\cite{BFSS}. 
Our next task is to clarify the correspondence for objects 
which are far from BPS. 
If there is no agreement between 
string theory and matrix model for such cases, 
we may need to look for a better (matrix) model.
\vfill
\section*{Appendix A}
\setcounter{equation}{0}
\renewcommand{\theequation}{A.\arabic{equation}}
In this appendix, 
we get commutation relations between modes' $\alpha_m$'s and $x$, 
namely, equations (\ref{amam})-(\ref{xx}).
The mode expansion for the coordinate shown in eq. (\ref{ModeEXP}) is  
\beq
X(\tau,\sigma)&=&x+M\bar{X}(\sigma^+) + \bar{X}(\sigma^-)\;, \n
\bar{X}(\rho)&:=& \sqrt{\alpha^\prime \over 2} \sum_{m \in {\bf Z}} 
 \int_0^{\rho}d\bar{\rho} e^{-i\left( m+iE \right)\bar{\rho}} \alpha_m\;. 
\label{mdexp}
\eeq
To use eq. (\ref{mdexp}) not only in the region $(0 \leq \sigma \leq \pi)$ but also in the region $(-\pi \leq \sigma \leq \pi)$,
we define extended variables in the following.
\beq
\cases{
\partial X_L(\rho) := M \partial X_R(\rho) & $\tau-\pi \leq \rho \leq \tau$ \cr
\partial X_R(\rho) := M^{-1} \partial X_L(\rho) & $\tau \leq \rho \leq \tau+\pi$ \cr}.
\eeq
Then we get useful commutation relations as follows.
\beq
\left[ \partial \bar{X}(\rho),\partial \bar{X}(\rho^\prime) \right]
= \pi i \alpha^\prime \delta^\prime(\rho-\rho^\prime)\eta 
\;\;,\;\;\;
\tau -\pi \leq \rho , \rho^\prime \leq \tau + \pi
\;.
\eeq
\subsection*{A.1 $\left[\alpha_m,\alpha_n \right]$}
Since we can write
\beq
\alpha_m = \frac{1}{2\pi} \sqrt{\frac{2}{\alpha^\prime}} \int_{\tau-\pi}^{\tau +\pi} d \rho e^{i \left( m + i E \right) \rho} \partial \bar{X}(\rho)\;,
\eeq 
we get the commutation relation between $\alpha_m$'s as follows.
\beq
\left[ \alpha_m , \alpha_n \right] 
&=& \frac{1}{(2\pi)^2}\frac{2}{\alpha^\prime} 
\int_{\tau-\pi}^{\tau +\pi} d \rho d \rho^\prime \left[ e^{i \left( m + i E \right) \rho} \partial \bar{X}(\rho), e^{i \left( n + i E \right) \rho \prime} \partial \bar{X}(\rho^\prime) \right] \n
&=&\frac{1}{(2\pi)^2}\frac{2}{\alpha^\prime} 
\int_{\tau-\pi}^{\tau +\pi} d \rho d \rho^\prime e^{i \left( m + i E \right) \rho} \left[ \partial \bar{X}(\rho), \partial \bar{X}(\rho^\prime) \right] \left(e^{i \left( n + i E \right) \rho \prime}\right)^t \n
&=&\frac{i}{2 \pi} \int_{\tau-\pi}^{\tau +\pi} d \rho d \rho^\prime 
\delta^\prime (\rho - \rho^\prime) e^{i \left( m+i E \right) \rho} \eta  
\left( e^{i \left( n+i E \right) \rho^\prime} \right)^t \n
&=&-\frac{i}{2 \pi} \int_{\tau-\pi}^{\tau +\pi} d \rho d \rho^\prime 
\delta (\rho - \rho^\prime) \left( \partial_\rho e^{i \left( m+i E \right) \rho} \right) \eta  
\left( e^{i \left( n+i E \right) \rho^\prime} \right)^t \n
&+& \frac{i}{2 \pi} \left. \int_{\tau-\pi}^{\tau +\pi} d \rho^\prime 
\delta (\rho - \rho^\prime) e^{i \left( m+i E \right) \rho} \eta  
\left( e^{i \left( n+i E \right) \rho^\prime} \right)^t \right|_{\rho = \tau -\pi}^{\rho =\tau +\pi} \;. \label{alal}
\eeq
Since we find that the last term in the eq. (\ref{alal}) is zero, 
the commutator is 
\beq
&&-\frac{i}{2 \pi} \int_{\tau-\pi}^{\tau +\pi} d \rho
\left( \partial_\rho e^{i \left( m+i E \right) \rho} \right) \eta  
\left( e^{i \left( n+i E \right) \rho^\prime} \right)^t \n
&&=\frac{m +i E}{2\pi} \int_{\tau-\pi}^{\tau +\pi} d \rho e^{i \left( m+i E \right) \rho } \eta \left(e^{i \left( n+i E \right) \rho } \right)^t 
\;. \label{alal2}
\eeq
The integrand in eq. (\ref{alal2}) is
\beq
e^{i(m+n) \rho } \eta\;,
\eeq
because of $M$ and $\bar{M}$ being $SO(D-1,1)$ matrices. 
Therefore we get
\beq
\left[\alpha_m , \alpha_n \right]=\delta_{m+n} \left(m + i E \right) \eta \;.
\eeq
\subsection*{A.2 $\left[x,\alpha_n \right]$}
We consider the following commutator
\beq
\left[\partial_\tau X(\tau,\sigma),X(\tau,\sigma^\prime)\right]
&=& \left[ \partial \bar{X}(\sigma^-), x \right] 
\n
&+& M\left[ \partial \bar{X}(\sigma^+),\bar{X}({\sigma^+}^\prime) \right] M^t 
+\left[ \partial \bar{X}(\sigma^-),\bar{X}({\sigma^-}^\prime) \right] 
\n
&+&M\left[ \partial \bar{X}(\sigma^+),\bar{X}({\sigma^-}^\prime) \right]
+\left[ \partial \bar{X}(\sigma^-),\bar{X}({\sigma^+}^\prime) \right]M^t
\;.\n
\eeq
Since
\beq
\left[ \partial \bar{X}(\rho),\bar{X}(\rho^\prime) \right]
&=&{\alpha^\prime \over 2} 
\sum_{m,n} e^{-i(m+i E)\rho}
\left[ \alpha_m,\alpha_n \right]
\int_0^{\rho^\prime} d\bar{\rho} \left(e^{-i(n+i E)\bar{\rho}}\right)^t 
\n
&=&-\pi i \alpha^\prime 
\left( \delta(\rho-\rho^\prime) - \delta(\rho) \right)\eta\;,
\eeq
we rewrite the commutator as
\beq
\left[\partial_\tau X(\tau,\sigma),X(\tau,\sigma^\prime)\right]
&=& \left[\partial_\tau X(\tau,\sigma) , x \right]\n
&-&2\pi i \alpha^\prime \delta(\sigma-\sigma^\prime)\eta 
-\pi i \alpha^\prime\left( M-M^{-1}\right)\delta(\sigma+\sigma^\prime)\eta  \n
&+&\pi i \alpha^\prime \Bigl[ 
\delta(\sigma^+)\left(1+M \right)
 + \delta(\sigma^-)\left(1+M^{-1}\right)
\Bigr]\eta\;.
\n
\eeq
Therefore
\beq
\left[\partial_\tau X(\tau,\sigma),X(\tau,\sigma^\prime)\right]
 &=& -2\pi i \alpha^\prime \delta(\sigma-\sigma^\prime)\eta \;,0 < \sigma,\sigma^\prime \le \pi 
\;,\\
&\Updownarrow& 
\n
\left[\partial_\tau X(\tau,\sigma) , x \right]
&=&-\pi i \alpha^\prime \Bigl[ 
\delta(\sigma^+)\left(1+M \right)
+\delta(\sigma^-)\left(1+M^{-1}\right)
\Bigr]\eta 
\;,\n
&\Updownarrow& 
\n
\left[x, \partial \bar{X}(\rho) \right]
&=&\pi i \alpha^\prime 
\delta(\rho)\left(1+M \right)\eta\;. \label{xdX}
\eeq
After integrating $\{$ $e^{i(m+iE)\rho}$ $\times$ eq. (\ref{xdX}) $\}$ 
by $\rho$ from $\tau-\pi$ to $\tau+\pi$, 
we get
\beq
\left[ x,\alpha_m \right] = i \sqrt{\alpha^\prime \over 2} \left(1+M \right)\eta\;.
\eeq 
\subsection*{A.3 $\left[x, x \right]$}
We consider the following commutator 
\beq
\left[X(\tau,\sigma),X(\tau,\sigma^\prime)\right]
&=&[x,x] \n
&+& M\left[ \bar{X}(\sigma^+),x \right] 
+\left[ \bar{X}(\sigma^-),x \right] \n
&+& \left[ x,\bar{X}({\sigma^-}^\prime) \right]
+\left[ x,\bar{X}({\sigma^+}^\prime) \right]M^t \n
&+& M\left[ \bar{X}(\sigma^+),\bar{X}({\sigma^+}^\prime) \right] M^t
+\left[ \bar{X}(\sigma^-),\bar{X}({\sigma^-}^\prime) \right] \n
&+& M\left[ \bar{X}(\sigma^+),\bar{X}({\sigma^-}^\prime) \right]
+\left[ \bar{X}(\sigma^-),\bar{X}({\sigma^+}^\prime) \right]M^t 
\;.\n
\eeq
We find
\beq
\left[ \bar{X}(\rho),\bar{X}(\rho^\prime) \right]
&=& {\alpha^\prime \over 2}\sum_{m,n} \int_0^\rho d\bar{\rho}\int_0^{\rho^\prime} d\bar{\rho^\prime} e^{-i(m+iE)\bar{\rho}} [\alpha_m,\alpha_n] \left( e^{-i(n+iE)\bar{\rho^\prime}}\right)^t \n
&=&\pi i \alpha^\prime \Big[ \varepsilon(\rho^\prime-\rho) 
+\varepsilon(\rho)- \varepsilon(\rho^\prime) \Big]\eta \;,
\\
\mbox{and \hspace{2cm}} && 
\n
\left[x,\bar{X}(\rho)\right] &=& \pi i \alpha^\prime \left(1+M \right)\varepsilon(\rho)\eta \;,
\eeq
where
\beq
\varepsilon(\rho)
:=\cases{
{1 \over 2} & $\rho >0$ \cr
0 & $\rho=0$ \cr
-{1 \over 2} & $\rho <0$ \cr
}\;. 
\eeq
Therefore we get
\beq
&
\left[ X(\tau,\sigma),X(\tau,\sigma^\prime) \right] =[x,x] -i \pi \alpha^\prime \left(M-M^{-1}\right) \varepsilon(\sigma+\sigma^\prime) \eta= 0  \;,0 < \sigma,\sigma^\prime \le \pi \;, 
&
\n
& \Updownarrow & \n 
&
[x,x]={\pi i \alpha^\prime \over 2}\left(M-M^{-1}\right)\eta\;.
& 
\eeq
\section*{Appendix B}
\setcounter{equation}{0}
\renewcommand{\theequation}{B.\arabic{equation}}
In this appendix, 
we rewrite the commutation relations between the modes 
in a more convenient form and apply the formula to our case.
Namely, we derive equations (\ref{baba})-(\ref{bxbx}).
We recall equations (\ref{amam})-(\ref{xx}):
\beq
\matrix{
\left[\alpha_m , \alpha_{-m} \right]&=& \left(m + i E \right) \eta 
\cr
\left[x,\alpha_m \right]&=&i\sqrt{\alpha^\prime \over 2}\left(1+M \right)\eta 
\cr
\left[x,x \right]&=&\frac{\pi i \alpha^\prime}{2}\left( M-M^{-1}\right)\eta 
\cr
}\;\;\;.
\label{comod}
\eeq
We would like to find the conjugate pairs for the modes. 
Our strategy is  
to transform the each right-hand side of equations (\ref{comod}) 
to become the following form, 
respectively.
\beq
\pmatrix{* & 0 \cr 0 & m \cr}\;\;,\;\;\;\pmatrix{0 & 0 \cr 0 & *\cr}\delta_m\;\;,\;\;\;
\pmatrix{* & 0 \cr 0 & 0\cr}\;.
\eeq
\subsection*{B.1 $T$}
First, we would like to diagonalize the matrix $E$ by transforming $\alpha_m$ by a matrix $T$
\beq
\breve{\alpha}_m &:=& T^{-1}\alpha_m
\;,\\
E_D &:=& T^{-1} E T =:\pmatrix{E_{11} & 0 \cr 0 & 0 \cr } 
\;,\\
\eta_T &:=& T^{-1} \eta T^{-t}\;,\;\; \mbox{Det}E_{11}\neq 0
\;. 
\eeq
Then the commutation relations are
\beq
\left[\breve{\alpha}_m , \breve{\alpha}_n \right]
&=& \left(m + i E_D \right) \eta_T \delta_{m+n}\;,  
\\
\left[x,\breve{\alpha}_m \right]
&=& i\sqrt{\alpha^\prime \over 2}\left(1+M \right)T \eta_T \;. \label{xalpha}
\eeq
\subsection*{B.2 $S$, $N$}
Second, we would like to write the each side of the equation (\ref{xalpha}) as the following form
by transforming $x$
\beq
\pmatrix{* & 0 \cr * & * \cr} 
\;,\n
y:=Sx \;. 
\eeq
How to determine the matrix $S$ and $N$ is as the following. 
\beq
S:=KJ\;\;,\;\;\;K:=\pmatrix{1 & -L_{12}L_{22}^{-1} \cr 0 & 1 \cr} 
\;,\\
L:=J\bar{L}\;\;,\;\;\;\bar{L}:=\left(1+M\right)T 
\;.
\eeq
$K$ is a matrix which transforms $L_{12}$ to $0$, 
if $L_{22}^{-1}$ exists.
$J$ is a matrix which exchanges the rows of a matrix ($\bar{L}$) 
so that $L_{22}^{-1}$ becomes existing 
\footnote{
The possibility is assumed here. 
If not so, 
we replace $L_{22}$ by its part whose minor is not zero. 
In our case, 
we avoid pure Dirichlet directions: $p+2 \cdots D-1$.}. 
That is, Det$L_{22} \neq 0$. \\
Then
\beq
N:=S\left(1+M\right)T=KL=\pmatrix{L_{11}-L_{12}L_{22}^{-1}L_{21} & 0 \cr L_{21} & L_{22}}=:\pmatrix{N_{11} & 0 \cr N_{21} & N_{22}}\;.
\n
\eeq
Then the commutation relations are
\beq
\left[y,\breve{\alpha}_m \right] 
&=& i \sqrt{\alpha^\prime \over 2} N \bar{\eta}  \label{ybalpha}  
\;,\\
\left[y,y \right] 
&=& {\pi i \alpha^\prime \over 2} \left( N \eta_{T} \left(S T\right)^t 
- \left(ST\right) \eta_T N^t \right)  \;. \label{yy}
\eeq
\subsection*{B.3 $C_m \cdots$ step $1$}
Third, we would like to write the each side of the equation (\ref{ybalpha}) as the following form by transforming $y$.
\beq
\pmatrix{0 & 0 \cr 0 & * \cr}\delta_m\;.
\eeq
We write $y$ as follows:
\beq
\breve{x}:=y-\sqrt{\alpha^\prime \over 2} \sum_{m \in {\bf Z}} C_m \breve{\alpha}_m\;.
\eeq
Then 
\beq
\left[\breve{x},\breve{\alpha}_m \right]
=\sqrt{\alpha^\prime \over 2} 
\Big[
i \pmatrix{N_{11} & 0 \cr N_{21} & N_{22} \cr} - C_{-m} \pmatrix{-m + i E_{11} & 0 \cr 0 & -m \cr }\eta_T
\Big]\;.
\eeq
Therefore we chose matrix $C_m$ as 
\beq
C_{m \neq 0}&:=&i N \left(m + i E_D\right)^{-1} 
\;,\n
C_0&:=&\pmatrix{N_{11}E_{11}^{-1} & c_{12} \cr N_{21}E_{11}^{-1} & c_{22} \cr} 
\;,
\eeq
where $c_{12}$ and $c_{22}$ will be defined in the next subsection.\\
Then we get
\beq
\left[\breve{x},\breve{\alpha}_m \right]
=i \sqrt{\alpha^\prime \over 2}\pmatrix{0 & 0 \cr 0 & N_{22}}\delta_m \;. 
\eeq
\subsection*{B.4 $C_m \cdots$ step $2$}
Finally, we would like to write the each side of the equation (\ref{yy}) as the following form by choosing $c_{12}$ and $c_{22}$.
\beq
\pmatrix{* & 0 \cr 0 & 0 \cr}\;.
\eeq
We can find commutation relation for $\breve{x}$'s as follows:
\beq
\left[\breve{x},\breve{x} \right]
&=&\left[y,y \right] \n
&+& {\pi i \alpha^\prime \over 2} \pmatrix{N_{11}\coth(\pi E_{11}) {\eta_T}_{11}N_{11}^t & N_{11} \coth(\pi E_{11}){\eta_T}_{11} N_{21}^t \cr N_{21} \coth(\pi E_{11}){\eta_T}_{11} N_{11}^t & N_{21}\coth(\pi E_{11}){\eta_T}_{11}N_{21}^t} \n
&+& {i\alpha^\prime \over 2}\pmatrix{0 &  c_{12}N_{22}^t \cr -N_{22}c_{12}^t & c_{22}N_{22}^t-N_{22}c_{22}^t }
\;,
\eeq
where we have used the formula
\beq
\sum_{m \in {\bf Z}}\frac{i}{m+iE_{11}} = \pi \coth{\pi E_{11}}\;.
\eeq
Therefore we choose $c_{12}$ and $c_{22}$ to satisfy following equations
\beq
c_{12}N_{22}^t&=&\pi N_{11}\coth(\pi E_{11}) {\eta_T}_{11} N_{21}^t 
+ \pi \left(N \eta_T (ST)^t -(ST) \eta_T N^t \right)_{12} 
\;,\n
c_{22}N_{22}^t - N_{22} c_{22}^t 
&=& \pi N_{21} \coth(\pi E_{11})
{\eta_T}_{11} N_{21}^t 
+ \pi \left(N \eta_T (ST)^t -(ST) \eta_T N^t \right)_{22} 
\;.\n
\eeq
Then we get the following formula:
\beq
\left[\breve{x},\breve{x} \right]={\pi i \alpha^\prime \over 2} 
\pmatrix{
{
N_{11} \coth{\pi E_{11}} {\eta_{T}}_{11} N_{11}^t 
\atop
+ N_{11}{\eta_{T}}_{11}{\left(S T \right)^t}_{11}- \left(S T \right)_{11} {\eta_{T}}_{11} N_{11}^t  
}
& 0 \cr 0 & 0 \cr
}
\;.
\eeq
\subsection*{B.5 conjugate momentum for $\breve{x}$}
We can define the conjugate momentum for the center of coordinate. 
\beq
\left[\breve{x},\breve{\alpha}_0 \right]=i \sqrt{\alpha^\prime \over 2}\pmatrix{0 & 0 \cr 0 & N_{22}}\;. \label{bxbalpha0}
\eeq
We assume that the determinant of the matrix $N_{22}$ is not zero. 
If not the case, 
we replace $N_{22}$ by its sub-matrix whose determinant is not zero.\\
We define the momenta $p^i$:
\beq
p^i:=\sqrt{2 \over \alpha^\prime}\left( N_{22}^{-t} \breve{\alpha}_0 \right)^i
\;,
\eeq
where the number of the independent momenta is determined by the rank of $N_{22}$.
$i$ is the index in the momentum space.
Then 
\begin{equation}
\left[\breve{x}^i, p^j\right]=i\delta^{ij}
\;.
\footnote{
If matrix $S$ is not $\eta$-othogonal matrix, 
we must find the momentum for $x$ not $\breve{x}$. 
}
\end{equation}
Now the trace of the zero mode in the phase space is defined as
\beq
\Tr_{p^i}:=\int \prod_{i} \frac{L_i}{2\pi} d p^i \;,
\eeq
where $L_i$ is a length of the area where $X^i$ covers~\cite{ACNY}.
\subsection*{B.6 two identical parallel D-branes case}
We can find as follows after straight forward calculation
\beq
\left[\breve{\alpha}_m, \breve{\alpha}_n\right] 
&=& \pmatrix{\matrix{0 & m+i \epsilon \cr m-i \epsilon  & 0 \cr}  & {\bf 0} \cr {\bf 0} & m{\bf 1} \cr}\delta_{m+n} 
\;,\;\;
\epsilon := \frac{1}{\pi} \ln\left( g +v \over g -v \right) \;,
\n
\left[\breve{x}^i,p^j\right] 
&=& i\delta^{ij}\;,\;\;
\left[\breve{x},\breve{\alpha}_{m \neq 0}\right] 
=\big[\breve{x},\breve{\alpha}_0^{\mu \neq i}\big]
=\left[\breve{x},\breve{x}\right]
= 0 
\;,\n
\breve{\alpha}_0^i 
&=:& \sqrt{2\alpha^\prime} \mbox{diag}\Big( \frac{1}{g}, {\bf 1} + m_2^t, \cdots ,{\bf 1} + m_{p+1 \over 2}^t \Big)^{ij} p^j 
\;,\n
g &:=& \sqrt{1-f_1^2(1-v^2)} \;,\;i,j=2 \cdots p+1 
\;,\n
\eeq
where $m_2, \cdots , m_{p+2 \over 2}$ are the $2\times2$ matrices which are defined in eq. (\ref{m_l}) 
and $f_1$ is a component of the field strength $F$ in eq. (\ref{F}).\\ 
$0$, $1$, $p+1$ components of matrices $T$, $S$, and $N$ are  
\beq
T&\sim&\pmatrix{
{1 \over \sqrt{2}} & {1 \over \sqrt{2}} & 0 \cr
{\sqrt{1-g^2} \over \sqrt{2}g } & -{\sqrt{1-g^2} \over \sqrt{2}g } 
& {1 \over g} \cr 
{1 \over \sqrt{2} g} & -{1 \over \sqrt{2} g} &  {\sqrt{1-g^2} \over g }\cr
}\;,\;\;
S\sim \pmatrix{1 & 0 & 0 \cr 0 & 0 & 1 \cr 0 & 1 & 0 \cr}\;, \n
N&\sim&\pmatrix{
{\sqrt{2}v \over g+v} & -{\sqrt{2}v \over g-v} & 0 \cr
{\sqrt{2} \over g+v} & -{\sqrt{2} \over g-v} & 0 \cr
{\sqrt{2} \sqrt{1-g^2}v \over g(g+v)} & {\sqrt{2}\sqrt{1-g^2}v \over g(g-v)} 
& {2 \over g} \cr
}
\;.
\eeq
Note that $\left[\breve{x},\breve{x}\right]$ is zero as in Ref~\cite{B}. 
\section*{Appendix C}
\setcounter{equation}{0}
\renewcommand{\theequation}{C.\arabic{equation}}
In this appendix we find energy momentum tensor which satisfies the ordinary operator product expansion (OPE).
Namely, we would like to explain eq. (\ref{EMT}) and eq. (\ref{TT}).
For this purpose, 
we start with the calculation of OPE between naive energy momentum tensor 
of the world sheet. 
We normal order it with respect to the vacuum defined in the section 2.
We define a naive energy momentum tenser as follows.
\beq
T_{a b}:=-4 \pi \alpha^\prime \frac{1}{\sqrt{h}}\frac{\delta S_{\rm string}}{\delta h^{a b}}\;, \label{defEMT}
\eeq
where $S_{\rm string}$ in (\ref{defEMT}) is that of eq. (\ref{Sstring}).\\
With 
\beq
z:=e^{i \rho}\;,
\eeq
we define the holomorphic energy momentum tensor:
\beq
T(z)
&:=&\frac{1}{2 \alpha^\prime z^2} \left(T_{00} + T_{01} \right) \n
&:=& \frac{1}{\alpha^\prime z^2} 
\left[
\partial_- X^\mu \partial_- X_\mu + \frac{i}{2} \psi_R^\mu \partial _- \psi_{R \mu} 
\right] + ({\rm ghost}) \n
&=&T_{\rm boson}(z)+ T_{\rm fermion}(z) + T_{\rm ghost}(z) \n
&:=& \frac{1}{2} \sum_{m,n \in {\bf Z}} z^{-m-n-2} :\alpha_m^T \eta \alpha_n:\n
&+&\frac{1}{2} \sum_{r,s \in {\bf Z} -\frac{1-a}{2}} z^{-r-s-2} :d_r^T \eta (s+iE) d_s: \n
&+& ({\rm ghost})\;.
\eeq
We extract the singular parts of the OPE in the following subsections.
\subsection*{C.1 OPE for bosonic part}
In this subsection, 
we calculate OPE between the bosonic part 
of the naive energy momentum tensor as follows:
\beq
T_{\rm boson}(z)T_{\rm boson}(w) 
&=& \frac{1}{2} \Tr_{\mu} \left \langle \partial Z(z) \partial Z(w) \right \rangle ^t \eta_T \left \langle \partial Z(z) \partial Z(w) \right \rangle  \eta_T 
\n
&+& \Tr_{\mu} \left \langle  \partial Z(z) \partial Z(w) \right \rangle ^t \eta_T :\partial Z(z) \partial Z(w) : \eta_T \n
&+& \mbox{regular( finite term with $w \rightarrow z$)}\;,
\eeq
where
\beq
T_{\rm boson}(z)&=&-\frac{1}{2}:\partial Z(z)^T \eta_T \partial Z(z): \n
&=&\frac{1}{2} \sum_{m,n}:\breve{\alpha}_n^t \eta_T \breve{\alpha}_m: z^{-m-n-2 }\;, \\
Z(z)&:=& \sqrt{\frac{2}{\alpha^\prime}} T^t \bar{X}(\rho=-i \ln z) \n
&=& \sum_{n \in {\bf Z}}z^{-n-i E_D} \breve{\alpha}_n\;.
\eeq
First, we calculate the two point function for $\partial Z$: 
\beq
\left \langle \partial Z(z)^\mu \partial Z(w)^\nu \right \rangle\;.
\eeq
We calculate this separately as follows.
\beq
\left \langle \partial Z(z)^\lambda \partial Z(w)^\delta \right \rangle
&=& -\sum_{m=1}^{\infty} \left( \frac{w}{z}\right)^m \frac{1}{z w} \left \langle \breve{\alpha}_m^\lambda \breve{\alpha}_{-m}^\delta \right \rangle \n
&=& -\frac{1}{(z-w)^2}\eta_T^{\lambda \delta} \;,\\
\left \langle \partial Z(z)^a \partial Z(w)^b \right \rangle
&=&-\frac{1}{z w} z^{-i \epsilon^a} w^{-i\epsilon^b} \left \langle \breve{\alpha}_0^a \breve{\alpha}_0^b \right \rangle \n
&-&\frac{1}{z w} \sum_{m=1}^{\infty} \left( \frac{w}{z}\right)^m z^{-i \epsilon^a} w^{-i\epsilon^b} \left \langle \breve{\alpha}_m^a \breve{\alpha}_{-m}^b \right \rangle \;, \label{dzdz}
\eeq
where $\lambda,\delta =2,\cdots, D-1$; 
$a, b=0, 1$ 
and $\epsilon^a:= (-)^a \epsilon$. 
\\
In the case of $(a,b)=(0,1)$, eq. (\ref{dzdz}) is
\beq
&&-\frac{1}{z w}\sum_{m=1}^{\infty}(m-i \epsilon) \left(\frac{w}{z}\right)^{m-i \epsilon} \eta_T^{01} \n
&=& -\frac{1}{z w} \left(\frac{w}{z}\right) \partial_{\frac{w}{z}}\sum_{m=1}^{\infty} \left(\frac{w}{z}\right)^{m-i \epsilon} \eta_T^{01} \n
&=& -\frac{1}{z^2}\left( \partial_{\frac{w}{z}} \frac{\left(\frac{w}{z}\right)^{i \epsilon}}{1-\frac{w}{z}}\right) \eta_T^{01}\;.
\eeq
In the case of $(a,b)=(1,0)$, 
eq. (\ref{dzdz}) is 
\beq
&& -\frac{1}{z w}\sum_{m=0}^{\infty}(m+i \epsilon) \left(\frac{w}{z}\right)^{m+i \epsilon} \eta_T^{10} 
\n 
&=& -\frac{1}{z w} \left(\frac{w}{z}\right) \partial_{\frac{w}{z}}\sum_{m=0}^{\infty} \left(\frac{w}{z}\right)^{m+i \epsilon} \eta_T^{10} \n
&=& -\frac{1}{w^2}\left( \partial_{\frac{z}{w}} \frac{\left(\frac{z}{w}\right)^{-i \epsilon}}{1-\frac{z}{w}}\right) \eta_T^{10}\;.
\eeq
That is, 
\beq 
\left \langle \partial Z(z)^\mu \partial Z(w)^\nu \right \rangle =
\pmatrix{ 
\matrix{ 
0 
&
-\frac{1}{w^2}\left( \partial_{\frac{z}{w}} \frac{\left(\frac{z}{w}\right)^{-i \epsilon}}{1-\frac{z}{w}}\right)  
\cr
-\frac{1}{z^2}\left( \partial_{\frac{w}{z}} \frac{\left(\frac{w}{z}\right)^{-i \epsilon}}{1-\frac{w}{z}}\right)
& 0  \cr
} 
 & {\bf 0} \cr 
{\bf 0} & -\frac{1}{(z-w)^2}{\bf 1} \cr 
}^{\mu \nu}. 
\n 
\eeq 
Therefore we get 
\beq 
T_{\rm boson}(z)T_{\rm boson}(w)&=&\frac{\frac{D}{2}}{(z-w)^4} 
+ \frac{2}{(z-w)^2}\left( T_{\rm boson}(w) - \frac{i\epsilon (1+i \epsilon)}{2 w^2}\right) \n 
&+& \frac{1}{z-w} \partial_w \left( T_{\rm boson}(w) - \frac{i\epsilon (1+i \epsilon)}{2 w^2} \right) + \mbox{regular}\;,
\n
\eeq 
where $D=10$ is the space-time dimension.
\subsection*{C.2 OPE for fermionic part}
We calculate OPE between the fermionic part of the energy momentum tensor:
\beq
T_{\rm fermion}=-{1 \over 2 \alpha^\prime z} : \psi_R \partial_z \psi_R:\;.
\eeq 
We can calculate the two point function for $\psi_R$ as follows.
\beq
\left \langle \psi_R(z)^\mu \psi_R(w)^\nu \right \rangle 
&=&  \alpha^\prime \sum _{r,s} \left \langle 
\left( z^{-\left( r+ iE \right)}  d_r \right)^\mu    
\left( w^{-\left( s+ iE \right)}  d_s \right)^\nu
\right \rangle \n
&=& \alpha^\prime \sum_{r,s} 
\left( z^{-\left( r + iE \right)} \right)^\mu_{\;\;\rho}
\left( w^{-\left( s + iE \right)} \right)^\nu_{\;\;\sigma}
\left \langle  d_r^\rho d_s^\sigma \right \rangle \n
&=& \alpha^\prime \sum_{s \leq 0 \atop r \geq 0 }
\left( z^{-\left( r + iE \right)} \right)^\mu_{\;\;\rho}
\left( w^{-\left( s + iE \right)} \right)^\nu_{\;\;\sigma}
\left( 
\eta^{\rho \sigma} \delta_{r+s} - \delta_r \delta_s \frac{1}{2} \eta^{\rho \sigma}
\right) \n
&=& \alpha^\prime \sum_{0 \leq r \in {\bf Z} -\frac{1-a}{2}} 
\left(
\left( z^{-\left( r + iE \right)} \right)
\eta  
\left( w^{-\left( -r + iE \right)} \right)^t 
\right)^{\mu \nu}
\left( 1-\frac{1}{2} \delta_r \right) \n
&=& \alpha^\prime \sum_{0 \leq r \in {\bf Z} -\frac{1-a}{2}}
\left(
z^{-\left( r + iE \right)} 
w^{-\left( -r - iE \right)} 
\eta 
\right)^{\mu \nu}
\left( 1-\frac{1}{2} \delta_r \right) \n
&=& \alpha^\prime \sum_{0 \leq r \in {\bf Z} +\frac{1-a}{2}}
\left( {w \over z} \right)^{r+iE}
\eta \left( 1-\frac{1}{2} \delta_r \right) \n
&=& \alpha^\prime \sum_{n=0}^{\infty}
\left( {w \over z} \right)^{n +{1-a \over 2} +iE}
\eta \left( 1-\frac{1}{2} \delta_{n + {1-a \over 2}} \right) \n
&=& \alpha^\prime 
\left(
\frac{1}{1-{w \over z}} - \frac{a}{2}
\right)
\left( \left( \frac{w}{z} \right)^{\frac{1-a}{2} + iE} \eta \right)^{\mu \nu} 
\;.
\eeq
Therefore, after straight forward calculation,
we get 
\beq
T_{\rm fermion}(z)T_{\rm fermion}(w)
&=& \frac{\frac{D}{4}}{(z-w)^4} \n
&+& \frac{2}{(z-w)^2}
\left[
T_{\rm fermion} + \frac{aD - 4 \Tr{E^2} }{16 w^2}
\right] \n
&+& \frac{1}{z-w} \partial_w 
\left[
T_{\rm fermion} + \frac{aD - 4 \Tr{E^2} }{16 w^2}
\right] \n
&+& ({\rm regular})\;.
\eeq
Note that $\Tr{E}=0$ is used in this calculation.
\subsection*{C.3 shifted energy momentum tensor}
Summarizing previous two subsections, using $\Tr{E^2}= 2\epsilon^2$, 
we define the  following shifted energy momentum tensor:
\beq
\widetilde{T(z)}&:=& T(z) - \frac{i\epsilon (1+i \epsilon)}{2 z^2} + \frac{aD - 8\epsilon^2}{16 z^2} \n
&=& T(z) + \frac{1}{z^2}\left( \frac{a D}{16} - \frac{i \epsilon}{2} \right)\;.
\eeq
We can get the following OPE.  
\beq
\widetilde{T(z)} \widetilde{T(w)} &=& \frac{\frac{3}{4} D}{(z-w)^4} + \frac{2 \widetilde{T(w)}}{(z-w)^2} + \frac{\partial_w \widetilde{T(w)}}{z-w} \n
&+& \mbox{regular} + \mbox{ ghost contribution }\;.
\eeq
This form is the usual type and is invariant under the exchange of the 
 $z$ and $w$ except for the regular part.
\vfill
\section*{Acknowledgments}
We would like to thank Y. Makeenko for useful comment 
to comparing imaginary parts of phase shifts.
\vfill


\begin{thebibliography}{99}
%
\bibitem{P}
J.~Polchinski, 
{\it Dirichlet branes and Ramond-Ramond Charges}, 
Phys. Rev. Lett. {\bf 75}(1995)4724-4727,
\href{http://xxx.lanl.gov/abs/hep-th/9510017}{hep-th/9510017}; 
J.~Polchinski, 
{\it TASI Lectures on D-Branes}, 
\href{http://xxx.lanl.gov/abs/hep-th/9611050}{hep-th/9611050}. 
%
\bibitem{IKKT}
N.~Ishibashi, H.~Kawai, Y.~Kitazawa and A.~Tsuchiya, 
{\it A Large--N Reduced Model as Superstring},   
Nucl. Phys. {\bf B498}(1997)467-491, 
\href{http://xxx.lanl.gov/abs/hep-th/9612115}{hep-th/9612115}.
%
\bibitem{M}
A.~Fayyazuddin, Y.~Makeenko, P.~Olesen, D.~J.~Smith, and K.~Zarembo, 
{\it Towards a non-Pertubative Formulation of IIB Superstring by Matrix Models}, Nucl. Phys. {\bf B499}(1997)159-182, 
\href{http://xxx.lanl.gov/abs/hep-th/9703038}{hep-th/9703038}. 
%
\bibitem{B}
C.~Bachas, 
{\it D-brane Dynamics}, 
Phys. Lett. {\bf B374}(1996)37-42, 
\href{http://xxx.lanl.gov/abs/hep-th/9511043}{hep-th/9511043}.
%
\bibitem{CT}
I.~Chepelev and A.~A.~Tseytlin, 
{\it Interaction of type IIB D-branes from D-instanton matrix model}, 
Nucl. Phys. {\bf B511}(1998)629-646, 
\href{http://xxx.lanl.gov/abs/hep-th/9705120}{hep-th/9705120}. 
%
\bibitem{MM}
B.~P.~Mandal and S.~Mukhopadhyay, 
{\it D-brane Interaction in the Type IIB Matrix Model}, 
Phys. Lett. {\bf B419}(1998)62-72, 
\href{http://xxx.lanl.gov/abs/hep-th/9709098}{hep-th/9709098}.
%
\bibitem{GG}
M.~B.~Green and M.~Gutperle, 
{\it Light-Cone Supersymmetry and D-branes}, 
Nucl. Phys. {\bf B476}(1996)484-514, 
\href{http://xxx.lanl.gov/abs/hep-th/9604091}{hep-th/9604091}.
%
\bibitem{L}
G.~Lifschytz, 
{\it Probing bound states of D-branes}, 
Nucl. Phys. {\bf B499}(1997)283-297, 
\href{http://xxx.lanl.gov/abs/hep-th/9610125}{hep-th/9610125}; 
G.~Lifschytz and  S.~D.~Mathur, 
{\it Supersymmetry and Membrane Interactions in M(atrix) Theory}, 
Nucl. Phys. {\bf B507}(1997)621-644, 
\href{http://xxx.lanl.gov/abs/hep-th/9612087}{hep-th/9612087}; 
G.~Lifschytz, 
{\it Four-Brane and Six-Brane Interactions in M(atrix) Theory},  
Nucl. Phys. {\bf B520}(1998)105-116, 
\href{http://xxx.lanl.gov/abs/hep-th/9612223}{hep-th/9612223}. 
\bibitem{ACNY}
A.~Abouelsaood, C.~G.~Callan, C.~R.~Nappi and S.~A.~Yost, 
{\it Open Strings in Background Gauge Fields }, 
Nucl. Phys. {\bf B280}(1987)599-624, 
%
\bibitem{BP} C.~Bachas and M.~Porrati, 
{\it Pair creation of open strings in an electric field },  
Phys. Lett. {\bf 296B}(1992)77-84, 
\href{http://xxx.lanl.gov/abs/hep-th/9209032}{hep-th/9209032}. 
%
\bibitem{K}
E.~Kiritsis, 
{\it Introduction To Superstring Theory}, 
\href{http://xxx.lanl.gov/abs/hep-th/9709062}{hep-th/9709062}. 
%
\bibitem{I}
N.~Ishibashi, 
{\it p-branes from (p-2)-branes in the Bosonic String Theory}, 
KEK-TH-570, 
\href{http://xxx.lanl.gov/abs/hep-th/9804163}{hep-th/9804163}. 
%
\bibitem{BFSS} 
T.~Banks, W.~Fischler, S.~H.~Shenker and L.~Susskind, 
{\it M Theory as a Matrix Model: a Conjecture}, 
Phys. Rev. {\bf D55}(1997)5112-5128, 
\href{http://xxx.lanl.gov/abs/hep-th/9610043}{hep-th/9610043}. 
%
%
\end{thebibliography}
\end{document}